\documentclass{IEEEtran}
\usepackage{amsmath}
\usepackage{amssymb}
\usepackage{amsthm}
\usepackage{dsfont}
\usepackage{etoolbox}
\usepackage{graphicx}
\usepackage{mathtools}
\usepackage{paralist}
\usepackage[caption=false,font=footnotesize]{subfig}
\usepackage[colorinlistoftodos]{todonotes}
\usepackage{url}
\usepackage{xspace}

\graphicspath{{./figures/}}

\theoremstyle{plain}
\newtheorem{prop}{Proposition}
\newtheorem{thm}{Theorem}
\newtheorem{cor}{Corollary}

\newtheorem{lem}{Lemma}

\theoremstyle{definition}

\newtheorem{exmp}{Example}

\theoremstyle{remark}
\newtheorem*{rem}{Remark}

\DeclareMathOperator{\head}{head}
\DeclareMathOperator{\tail}{tail}
\DeclareMathOperator{\ext}{Ext}

\newcommand*{\fig}[1]{\figurename~\ref{fig:#1}\xspace}
\newcommand*{\eq}[1]{\eqref{eq:#1}\xspace}
\newcommand*{\lemref}[1]{Lemma~\ref{lem:#1}\xspace}
\newcommand*{\thmref}[1]{Theorem~\ref{thm:#1}\xspace}
\newcommand*{\propref}[1]{Proposition~\ref{prop:#1}\xspace}
\newcommand*{\exmpref}[1]{Example~\ref{exmp:#1}\xspace}
\newcommand*{\appendixproof}[1]{See Appendix~\ref{sec:#1}\xspace}
\newcommand*{\secref}[1]{\S\ref{sec:#1}\xspace}

\newtoggle{proofs}
\togglefalse{proofs}
\toggletrue{proofs}

\newtoggle{extras}
\togglefalse{extras}

\title{Structural and Optimization Properties for Joint Selection of Source Rates and Network Flow}
\author{%
	Bradford~D.~Boyle,~\IEEEmembership{Student~Member,~IEEE,} and %
	Steven Weber,~\IEEEmembership{Senior~Member, IEEE}%
	\thanks{%
		This work was partially funded by the National Science Foundation Award \#1228847 and the Air Force Research Laboratory under agreement number FA9550-12-1-0086. %
		The U.S. Government is authorized to reproduce and distribute reprints for Governmental purposes notwithstanding any copyright notation thereon.%
	}%
	\thanks{%
		B.~Boyle and S.~Weber are with the Department of Electrical and Computer Engineering, Drexel University, Philadelphia, PA USA (email: \textsf{bradford@drexel.edu} and \textsf{sweber@coe.drexel.edu}).
		Preliminary results were presented at the Data Compression Conference (DCC), 2014 \cite{BoyWeb2014}.
	}%
}

\begin{document}
\maketitle
\begin{abstract}
	We consider the optimal transmission of distributed correlated discrete memoryless sources across a network with capacity constraints.
	We present several previously undiscussed structural properties of the set of feasible rates and transmission schemes.
	We extend previous results concerning the intersection of polymatroids and contrapolymatroids to characterize when all of the vertices of the Slepian-Wolf rate region are feasible for the capacity constrained network.
	An explicit relationship between the conditional independence relationships of the distributed sources and the number of vertices for the Slepian-Wolf rate region are given.
	These properties are then applied to characterize the optimal transmission rate and scheme and its connection to the corner points of the Slepian-Wolf rate region.
	In particular, we demonstrate that when the per-source compression costs are in tension with the per-link flow costs the optimal flow/rate point need \emph{not} coincide with a vertex of the Slepian-Wolf rate region.
	Finally, we connect results for the single-sink problem to the multi-sink problem by extending structural insights and developing upper and lower bounds on the optimal cost of the multi-sink problem.
\end{abstract}
\begin{IEEEkeywords}
	Distributed source coding, minimum cost network flow, linear programming
\end{IEEEkeywords}
\section{Introduction}
\label{sec:introduction}
\subsection{Motivation}
A class of problems that arise in many contexts is the transmission of distributed discrete memoryless sources across a capacity-constrained network to a collection of sinks.
Information theoretic characterizations of this class of problems has received much attention in recent years as a result of the development of network coding \cite{AhlCaiLi2000} and can be traced back to the seminal work of Slepian and Wolf \cite{SleWol1973}.
In this paper, we consider the design problem of selecting a set of rates and a transmission scheme for a given network that are optimal with respect to known information-theoretic characterizations.
A necessary assumption is that all sinks want all sources.
The general case where each sink wishes to receive a subset of the sources has an implicit characterization in terms of the region of entropic vectors and only inner and outer bounds are explicitly known \cite{SonYeu2001,Yeu2008}.
\begin{figure}
	\centering
	\includegraphics{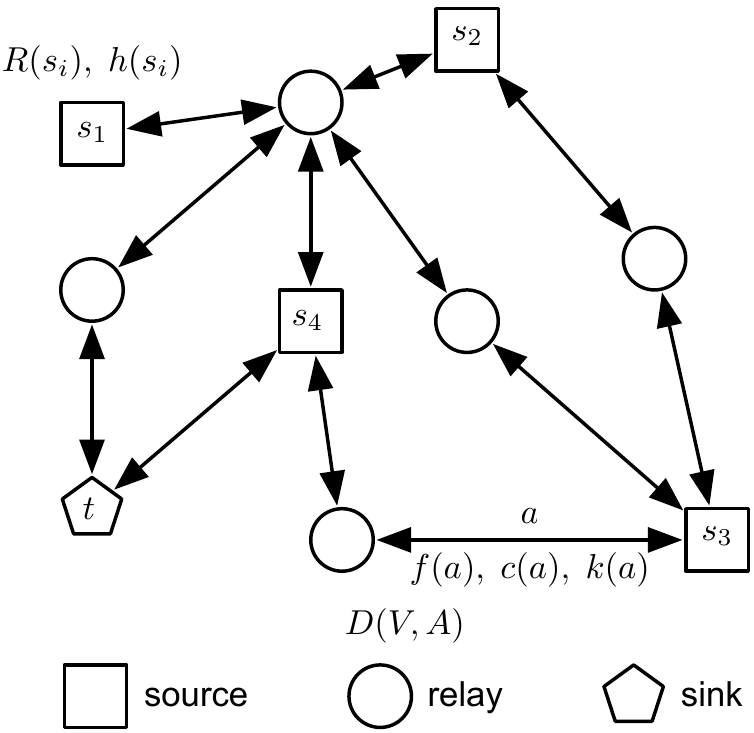}
	\caption{Problem overview: A single sink node losslessly recovers distributed correlated sources over a capacity constrained network. There is a per unit cost associated with activating each link as well as a per unit cost at the sources.}
	\label{fig:slepian-wolf-overview}
\end{figure}

\subsection{Related Work}
\label{sec:related-work}
Han considers the problem of communicating a distributed set of correlated sources to a single sink across a capacity-constrained network and characterizes the set of achievable rates \cite{Han1980}.
For a single sink, it is known that the min-cut/max-flow bounds can be achieved \cite{Yeu2008} and in particular, Slepian-Wolf (SW) style source coding \cite{SleWol1973} followed by routing is sufficient \cite{Han1980}.
Han proposes a minimum-cost problem where link activations are charged a per unit cost and cites work by Fujishige \cite{Fuj1978} as an algorithmic solution to the proposed problem.
The proposed algorithm can be applied to problems with both link and source costs; however, it cannot be extended to the case of multiple sinks.
Additionally, the algorithm is only guaranteed to terminate in finite time if the data are assumed integral \cite{Fuj1978}.
Barros et al.\ \cite{BarSer2006} contains a similar characterization of the set of achievable rates and an identical LP formulation as \cite{Han1980} but no discussion of an efficient algorithm.
In the achievability proof of Barros et al. (and Han \cite{Han1980}), a separation between the source encoder rates and the network flows is observed, leading to a natural mapping of this problem into the traditional protocol stack.

When the problem is extended from a single sink to multiple sinks, each sink required to receive all the sources, it is known that
\begin{inparaenum}[i)]
\item in general routing is not sufficient for achieving the min-cut/max-flow bounds;
\item network coding is necessary \cite{AhlCaiLi2000}, and;
\item in fact linear network coding is sufficient \cite{Yeu2008}.
\end{inparaenum}
Identical characterizations of when a distributed correlated source can be multicast across a capacity-constrained network have been given by Song et al., Ramamoorthy, and Han \cite{SonYeu2001,Ram2011,Han2011}.
These characterizations are a natural extension of the result for a single sink \cite{Han1980}.
Earlier work by Cristescu et al.\ also considers the problem of SW coding across a network with links that were \emph{not} capacity-constrained \cite{CriBefVet2005}.
This allows for an optimal solution to be obtained as the superposition of minimum weight spanning trees.
Two key differences between the work of Ramamoorthy \cite{Ram2011} and Han \cite{Han2011} are that the former makes the assumption of rational capacities to make use of results from \cite{HoMedKoe2006} and specifically considers the problem of minimizing the cost to multicast the sources.
Focusing on lossless communication and assuming a linear objective, the cost to multicast the sources can be formulated as a linear objective with per unit cost for activating links.
By not having a per-source cost, the proposed LP can be solved by applying dual decomposition to exploit the combinatorial structure of the SW rate region associated with the correlated sources and using the subgradient method to approximate the optimal cost \cite{Ram2011}.
In the present work, we consider a more general model by including a per-unit rate cost for each source node.
The technique of dual decomposition and application of the subgradient method has been used in work by Yu et al.\ \cite{YuYua2005} and Lun et al.\ \cite{LunRatMed2006}.
Yu et al.\ considers the problem of \emph{lossy} communication of a set of sources and minimizes a cost function that trades off between the estimation distortion and the transmit power of the nodes in the network.
The rate-distortion region is, in general, not polyhedral and the resulting optimization problem is convex.
Lun et al.\ makes the assumption of a single source and therefore does not deal with the interdependencies among the different source rates.

\subsection{Summary of Contributions}
\label{sec:summary}
Previous works have only considered the dual with respect to a subset of the constraints in order to exploit the contrapolymatroidal structure of the SW rate region.
In the present work, we restrict our attention to a single sink and more fully investigate the underlying combinatorial structure of the resulting set of achievable rates.
By considering the full dual LP, we demonstrate the application of the additional structural properties towards the development of alternative algorithmic solutions.

The rest of the paper is organized as follows:
In \secref{preliminaries}, we present and discuss relevant supporting material from literature as well as formally pose our optimization problem.
In \secref{feasible-set}, we extend existing results concerning the intersection of polymatroid with a contrapolymatroid and characterize their types of intersections.
We also relate the conditional independence relationships of the sources to degeneracy of the extreme points of the Slepian-Wolf rate region, reducing the number of inequality constraints needed to describe the polyhedron.
In \secref{sufficient-conditions}, we consider the dual of the linear program to develop sufficient conditions for optimal solutions.
We are particularly interested in knowing when the optimal solution will coincide with a vertex of the Slepian-Wolf rate region.
We demonstrate that when there is an imbalance between the source costs and flow costs (i.e., cheap compression and expensive routing vs.\ expensive compression and cheap routing), the optimal rate allocation may not coincide with a vertex of the rate region.
In \secref{multiple-sinks}, we partially extend our results to the multi-sink problem and bound the optimal value of the multi-sink problem with the optimal values of related single sink problems.
We conclude in \secref{conclusion}.
\section{Preliminaries}
\label{sec:preliminaries}
We model the network as a simple directed graph \(D = (V,A)\) with nodes \(V\) representing alternately sources, routers, and destinations, and arcs \(A\) representing network connections between nodes in \(V\).
We model the arcs \(A\) as capacitated with capacity \(c = (c(a), a \in A)\).
If \(a = (u,v) \in A\), then we define \(\tail(a) \triangleq u\) and \(\head(a) \triangleq v\) and
\begin{subequations}
	\begin{gather}
		\delta^{out}(v) \triangleq \left\{a \in A: \tail(a) = v\right\}\\
		\delta^{in}(v) \triangleq  \left\{a \in A: \head(a) = v\right\}.
	\end{gather}
\end{subequations}
For an arbitrary set function \(f: U \mapsto \mathds{R}\), we denote \(\sum_{u \in B} f(u)\) by \(f(B)\) for any subset \(B \subseteq U\).

The distributed sources are located at a subset \(S \subset V\) of the network elements and need to be collected at a sink \(t \in V \setminus S\).
We model the sources as a collection of correlated discrete memoryless random variables \((X_s : s \in S)\).
There is a joint distribution \(p_{(X_s : s \in S)}\) (shortened to just \(p_S\)) on the set of sources which in turn gives rise to a vector of conditional entropies \((H(X_{U}|X_{U^\mathsf{c}}), U \subseteq S)\), where \(H(X_{U}|X_{U^\mathsf{c}})\) is the conditional entropy associated with the subset of sources \(U \subseteq S\) given the values of the other sources \(U^\mathsf{c} = S \setminus U\).

The decision variables in our model are both
\begin{inparaenum}[i)]
\item the rates for each source, \(R = (R(s), s \in S)\), and
\item the flow on each arc, \(f = (f(a), a \in A)\).
\end{inparaenum}
The rate \(R(s)\) is the rate at which source \(s\) transmits, which must be routed (possibly split over multiple paths) towards the destination \(t\), and the flow \(f(a)\) is the superposition over all rates \(R(s)\) whose routes traverse arc \(a\).
Flows must satisfy:
\begin{inparaenum}[i)]
\item capacity constraints (\(0 \leq f(a) \leq c(a)\) for all \(a \in A\)), and
\item conservation of flow at all non-source, non-sink nodes (\(f(\delta^{out}(v)) = f(\delta^{in}(v))\) for all \(v \in V \setminus (S \cup \{t\})\)).
\end{inparaenum}
A flow \(f\) supports rates \(R\) if for all \(s \in S\), \(R(s) = f(\delta^{out}(s)) - f(\delta^{in}(s))\).
The novelty of our optimization problem model lies in jointly optimizing over both \((f,R)\) simultaneously, since most of the network flow literature assumes the source rates to be an input to the flow problem.
While the multi-source network coding problem includes variables for both source rates and edge rates (analogous to our flow variables), much of the network coding literature has focused on characterizing the region obtained by projecting onto either the source rate or edge rate variables.
Our work focuses on the cases where rate regions are known and expressly considers the problem of joint optimization \emph{without} the projection onto one set of variables.
For the case of multiple sinks, routing will no longer be sufficient and we will need to consider network coding.
In this case, there will be a ``virtual'' flow \(f_t\) for each sink \(t\) satisfying the normal flow constraints.
Under network coding, the physical flow \(f(a)\) on an arc \(a\) will then satisfy \(f_t(a) \leq f(a)\) for all \(t\) \cite{LunRatMed2006}.

We begin with the Slepian-Wolf theorem, which characterizes the set of source rates for which lossless distributed source codes exist.
\begin{thm}[Slepian-Wolf \cite{SleWol1973}]
	\label{thm:slepian-wolf}
	The rate region \(\mathcal{R}_{SW}\) for distributed lossless source coding the discrete memoryless sources \(X_S\) is the set of rate tuples \(R\) such that
	\begin{equation}
		R(U) \geq H(X_{U} | X_{S \setminus U}) \quad \forall \; U \subseteq S.
	\end{equation}
\end{thm}
For brevity, let us define \(\sigma_{SW}: 2^{|S|} \rightarrow \mathds{R}\) as
\begin{equation}
	\label{eq:conditional-entropy}
	\sigma_{SW}(U) \triangleq H(X_{U} | X_{U^\mathsf{c}})
\end{equation}
which is a nonnegative, nondecreasing supermodular set function on the set of sources.
Note that the rate region of \thmref{slepian-wolf} is the contrapolymatroid \(Q_{\sigma_{SW}}\) associated with \(\sigma_{SW}\):
\begin{equation}
	\mathcal{R}_{SW} = Q_{\sigma_{SW}} \triangleq \left\{R \in \mathds{R}^{|S|} : R(U) \geq \sigma_{SW}(U), \; \forall \; U \subseteq S\right\}.
\end{equation}

The following theorem characterizes the set of source rates for which there exists a supporting flow.
\begin{thm}[Megiddo \cite{Meg1974}]
	\label{thm:meggido}
	There exists a flow \(f\) that supports the rates \(R\) iff
	\begin{equation}
		R(U) \leq \min\{c(\delta^{out}(X)) : U \subseteq X, t \in V \setminus X\} \quad \forall \; U \subseteq S.
	\end{equation}
\end{thm}
Paralleling \eq{conditional-entropy}, define \(\rho_c: 2^{|S|} \rightarrow \mathds{R}\) as
\begin{equation}
	\rho_c(U) = \min\{c(\delta^{out}(X)) : U \subseteq X, t \in V \setminus X\}
\end{equation}
This is the min-cut capacity/max-flow value from the set \(U\) to the sink \(t\), which is a nonnegative, nondecreasing submodular set function on the set of sources.
The set of source rates for which there exists a supporting flow is the polymatroid \(P_{\rho_c}\) associated with \(\rho_c\):
\begin{equation}
	P_{\rho_c} \triangleq \left\{R \in \mathds{R}^{|S|} : R \geq 0, R(U) \leq \rho_c(U), \; \forall \; U \subseteq S\right\}.
\end{equation}

\begin{figure}
	\centering
	\includegraphics{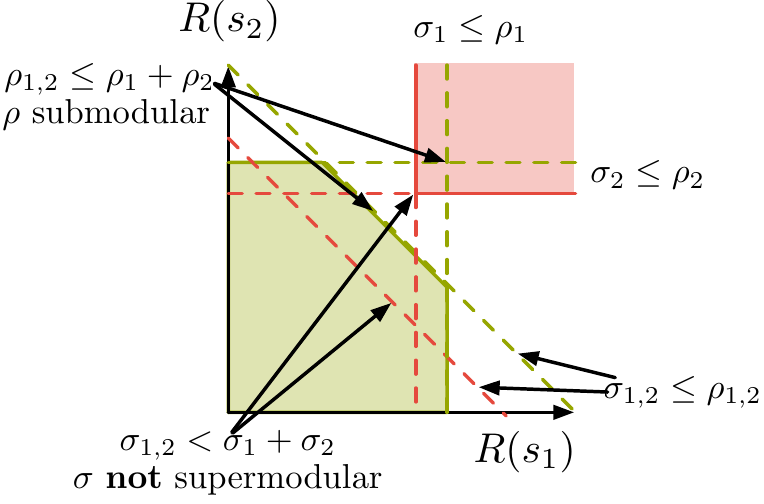}
	\caption{%
		An example of set functions \(\rho\) and \(\sigma\) that satisfy \(\sigma(U) \leq \rho(U)\) for all \(U \subseteq S\). %
		We see that \(\rho\) is submodular since \(\rho(\{s_1, s_2\}) \leq \rho(\{s_1\}) + \rho(\{s_2\})\), while \(\sigma\) is \emph{not} supermodular since \(\sigma(\{s_1, s_2\}) < \sigma(\{s_1\}) + \sigma(\{s_2\})\). %
		In this case, \thmref{han-matching} cannot be used to conclude that \(P_\rho\) and \(Q_\sigma\) have a non-empty intersection.%
	}
	\label{fig:sufficiency-example}
\end{figure}

The final theorem in this section characterizes when the intersection of the sets of source rates from the previous two theorems is non-empty.
\begin{thm}[Han's matching condition \cite{Han1980}]
	\label{thm:han-matching}
	Let \(\sigma\) and \(\rho\) be supermodular and submodular set functions, respectively.
	Then
	\begin{equation}
		I_{\sigma, \rho} \triangleq Q_\sigma \cap P_\rho \neq \varnothing
	\end{equation}
	if and only if
	\begin{equation}
		\label{eq:han-matching}
		\sigma(U) \leq \rho(U) \quad U \subseteq S
	\end{equation}
	In particular, there exists distributed lossless source codes for communicating the sources \(X_S\) across the capacity-constrained network to the sink \(t\) iff \(\sigma_{SW}(U) \leq \rho_c(U)\) for all \(U \subseteq S\).
\end{thm}
As mentioned in \cite{Han1980}, the proof of the necessity of \thmref{han-matching} is obvious.
The proof of the sufficiency of \thmref{han-matching} depends critically on the submodularity of \(\rho\) and supermodularity of \(\sigma\).
\fig{sufficiency-example} gives an example of generic set functions \(\sigma\) and \(\rho\) that satisfy \eq{han-matching} for which the set \(I_{\sigma, \rho} = \varnothing\) because \(\sigma\) is \emph{not} supermodular.
Specializing \eq{han-matching} to conditional entropy and min-cut capacity gives
\begin{equation}
	H(X_U \mid X_{U^\mathsf{c}}) \leq \rho_c(U)
\end{equation}
which has the following interpretation: \(H(X_U \mid X_{U^\mathsf{c}})\) is the information only available at the set of sources \(U\) and the network must be able to at least support a flow of that value from those sources.

Our objective is to route the information from the sources \(S\) to the sink \(t\) as efficiently as possible, which we measure via costs on both the rate of the sources, and the costs of activating the arcs.
Specifically, let \(h = (h(s), s \in S)\) be the cost per bit per second associated with each source, and \(k = (k(a), a \in A)\) be the cost per unit flow associated with each arc.

With this notation, the cost of a solution \((f,R)\) is \(k^\intercal f + h^\intercal R\).
The constraints are the natural ones given the model description above:
\begin{inparaenum}[i)]
\item flows must observe the arc capacity constraints \(f \leq c\),
\item flows \(f\) and rates \(R\) must satisfy conservation of flow at all router nodes \(v \in V \setminus (S \cup t)\),
\item the flows and rates must match at the sources, so that the inflow plus the source rate equals the outflow, and
\item the rates must be large enough to fully describe the source entropies \(R(U) \geq H(X_U|X_{U^\mathsf{c}})\) for all \(U \subseteq S\).
\end{inparaenum}
By only considering a single sink, we only need to find one flow vector \(f\).
For the general network coding case, the model can be extended in a natural way to account for the ``virtual'' flow for each sink and the physical flow on each arc.

The linear program described above is as follows:
\begin{equation}
	\label{eq:primal}
	\begin{aligned}
		& \underset{f \geq 0, R}{\text{minimize}} & & \sum_{a \in A}k(a)f(a) + \sum_{s \in S}h(s)R(s) & &\\
		& \text{subject to} & & f(a) \leq c(a) & & a \in A\\
		& & & f(\delta^{in}(v)) - f(\delta^{out}(v)) = 0 & &v \in N\\
		& & & R(s) + f(\delta^{in}(s)) - f(\delta^{out}(s)) = 0 & & s \in S\\
		& & & R(U) \geq H(X_U | X_{U^\mathsf{c}}) & &U \subseteq S
	\end{aligned}
\end{equation}
where \(N \triangleq V \setminus (S \cup \{t\})\), \(f(\delta(v)) \triangleq \sum_{a \in \delta(v)}f(a)\), and \(R(U) \triangleq \sum_{s \in U}R(s)\), \(U \subseteq S\).
The linear program in \eq{primal} has \(|A| + |V| - 1 + 2^{|S|}\) inequalities.
If \(|S| = \mathcal{O}(|V|)\), then the LP is exponential in the size of the graph.
Observe that an optimal solution \((f^*, R^*)\) to \eq{primal} will satisfy \(R^*(S) = H(X_S)\) \cite{Han1980}.

\section{Feasible Set Structural Properties}
\label{sec:feasible-set}
We see from \thmref{slepian-wolf} and \thmref{meggido} that the set of feasible rates \(Q_{\sigma_{SW}} \cap P_{\rho_c}\) is the intersection of a polymatroid with a contrapolymatroid.
The resulting polytope can be thought of as being obtained by the projection \(p: \mathds{R}^{|A|+|S|} \rightarrow \mathds{R}^{|S|}\) of the set of feasible \((f, R)\) tuples onto the rate variables \(R\).
In this section we present several structural properties of the set of feasible \((f, R)\) and the associated lower dimensional set \(Q_{\sigma_{SW}} \cap P_{\rho_c}\) that are \emph{independent} of the assumed objective function in \eq{primal}.

\subsection{General properties from sub-/supermodularity}
For any polyhedron \(P\), we denote the set of extreme points as \(\ext(P)\).
The extreme points (vertices) of a contrapolymatroid \(Q_\sigma\) are given by
\begin{equation}
	R_\pi(s_{\pi(i)}) = \sigma(U_{\pi(i)}) - \sigma(U_{\pi(i-1)}) \quad i = 1, \ldots, |S|
\end{equation}
where \(\pi\) ranges over all permutations of \([|S|]\)\footnote{For an integer \(i\), the set \(\{1, \ldots, i\}\) is denoted by \([i]\).} and \(U_{\pi(i)} = \{s_{\pi(1)}, \ldots, s_{\pi(i)}\}\) \cite{Sch2003b}.
The extreme rays of \(Q_\sigma\) are the unit vectors of \(\mathds{R}^{|S|}\).
Similarly, the extreme points of a polymatroid \(P_\rho\) are given by
\begin{equation}
	\label{eq:polymatroid-vertex}
	R_\pi(s_{\pi(i)}) =
	\begin{dcases}
		\rho(U_{\pi(i)}) - \rho(U_{\pi(i-1)}) & i \leq k\\
		0 & i > k
	\end{dcases}
\end{equation}
where \(\pi\) ranges over all permutations of \([|S|]\) and where \(k\) ranges over \(0,\ldots,|S|\) \cite{Sch2003b}.
With these definitions, we can now show that the half-space inequalities for \(U_{\pi(i)}\) hold with equality.
\begin{lem}
	\label{lem:vertex-sum-chain}
	If \(R_\pi\) is the vertex of \(Q_\sigma\) corresponding to permutation \(\pi\) then
	\begin{equation}
		R_\pi(U_{\pi(i)}) = \sigma(U_{\pi(i)}).
	\end{equation}
	If \(R_\pi\) is the vertex of \(P_\rho\) corresponding to permutation \(\pi\) then
	\begin{equation}
		R_\pi(U_{\pi(i)}) = \rho(U_{\pi(i)}).
	\end{equation}
\end{lem}
\iftoggle{proofs}{\begin{IEEEproof}\appendixproof{vertex-sum-chain-proof}\end{IEEEproof}}{}
The \emph{base polyhedron} of \(Q_\sigma\) and \(P_\rho\) is defined as \cite{Fuj2005}
\begin{subequations}
	\begin{gather}
		B_\sigma \triangleq Q_\sigma \cap \{R : R(S) = \sigma(S)\}\\
		B_\rho \triangleq P_\rho \cap \{R : R(S) = \rho(S)\}.
	\end{gather}
\end{subequations}
As noted previously, an optimal solution \((f^*, R^*)\) to the LP \eq{primal} will satisfy \(R^*(S) = H(X_S)\) and thus \(R^* \in B_{\sigma_{SW}}\).

In general, Han's matching condition (\thmref{han-matching}) does not allow us to conclude if the base polyhedron of a contrapolymatroid \(B_\sigma\) is wholly contained in the intersection \(Q_{\sigma} \cap P_{\rho}\).
\begin{exmp}
Consider \(S = \{s_1, s_2\}\) and let \(\rho\) be submodular and \(\sigma\) supermodular such that \(\sigma(U) \leq \rho(U)\) for all \(U \subseteq S\).
Consider the vertex \(R = (\sigma(s_1), \sigma(s_1, s_2) - \sigma(s_1))\) of \(Q_\sigma\).
We have, by the assumption of \eq{han-matching} that \(R(s_1) = \sigma(s_1) \leq \rho(s_1)\) and \(R(s_1) + R(s_2) = \sigma(s_1, s_2) \leq \rho(s_1, s_2)\).
From the supermodularity of \(\sigma\), we have that \(\sigma(s_2) \leq \sigma(s_1, s_2) - \sigma(s_1)\) and by assumption \(\sigma(s_2) \leq \rho(s_2)\); this \emph{does not} allow us to conclude one way or the other if \(\sigma(s_1, s_2) - \sigma(s_1) \gtrless \rho(s_2)\) and so we cannot, in general, determine if \(R \in P_\rho\) and therefore \(R \in I_{\sigma, \rho}\).
\qed
\end{exmp}

\begin{figure}
	\centering
	\subfloat[]{\includegraphics[width=156.0pt]{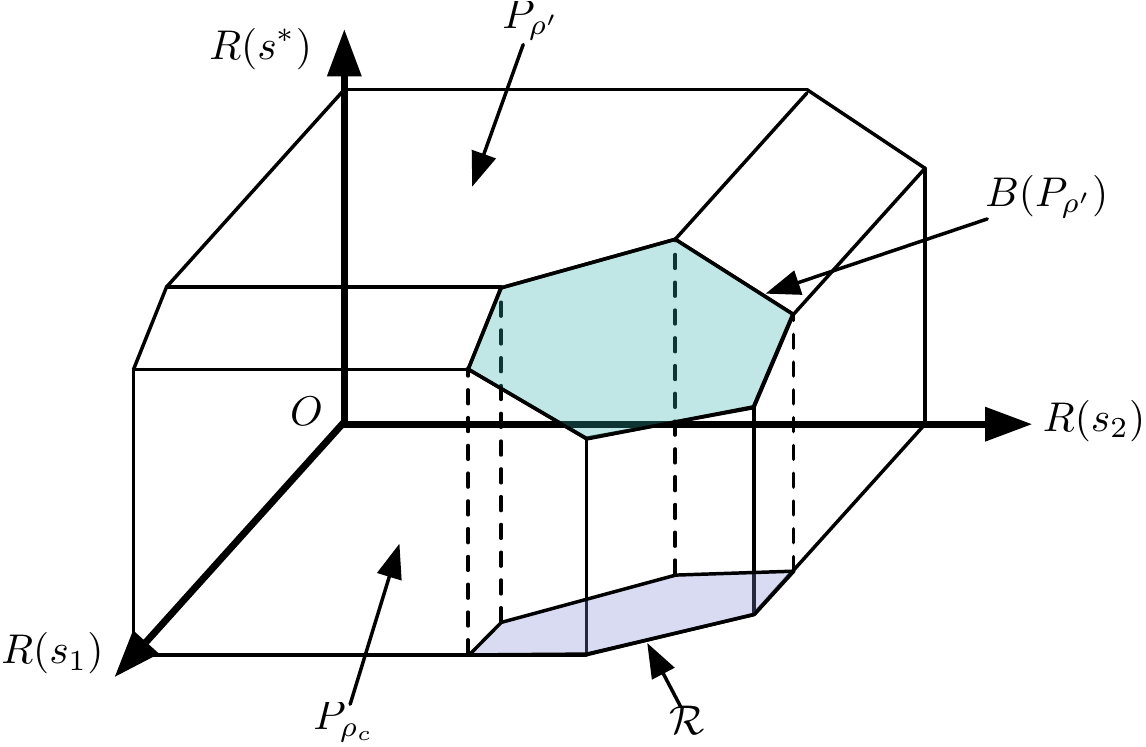}\label{fig:polymatroid-projection}}
	\subfloat[]{\includegraphics{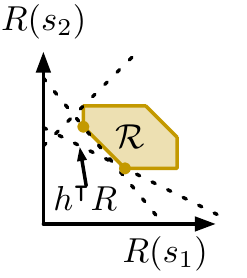}\label{fig:generalized-polymatroid}}
	\caption{Generalized polymatroid as the projection of the base polytope \(B_f\) of a polymatroid \(P_f\) in a higher dimension. %
	Minimization of a linear objective with sign-definite weight vector \(h\) will have a solution at a vertex of \(B_\sigma\) (\(h \geq 0\)) or at a vertex of \(B_\rho\) (\(h \leq 0\)).}
\end{figure}

Our first set of results characterize when \(B_\sigma\) and \(B_\rho\) are contained in \(I_{\sigma, \rho}\).
For generic submodular \(\rho\) and supermodular \(\sigma\) set functions we assume, w.l.o.g., that \(\sigma(\varnothing) = \rho(\varnothing) = 0\).
We begin by combining results from Frank et al. \cite{FraTar1988} and Fujishige \cite{Fuj1984,Fuj2005} and provide an explicit characterization of the vertices of \(I_{\sigma, \rho}\) for certain instances of \(\sigma\) and \(\rho\).
\begin{thm}
	\label{thm:intersection}
	Let \(\sigma\) be a supermodular set function and \(\rho\) be a submodular set function.
	If
	\begin{equation}
		\label{eq:intersection-thm}
		\sigma(U) - \sigma(U \setminus T) \leq \rho(T) - \rho(T \setminus U) \quad \forall \; T, U \subseteq S
	\end{equation}
	then the vertices of \(I_{\sigma, \rho}\) are given by
	\begin{equation}
		\label{eq:generalized-polymatroid-vertices}
		R_\pi^j(s_{\pi(i)}) =
		\begin{dcases}
			\rho(U_{\pi(i)}) - \rho(U_{\pi(i-1)}) & i \leq j\\
			\sigma(S \setminus U_{\pi(i-1)}) - \sigma(S \setminus U_{\pi(i)}) & i > j
		\end{dcases}
	\end{equation}
	where \(\pi\) is a permutation and \(j \in \{0, \ldots, n\}\).
\end{thm}
\begin{IEEEproof}
	\iftoggle{proofs}{\appendixproof{intersection-proof}}{Omitted for brevity.}
\end{IEEEproof}
Ignoring the flow costs in the LP of \eq{primal}, we see that an optimal solution corresponds to an extreme point of \(Q_{\sigma_{SW}}\).
\begin{cor}
	\label{cor:generalized-polymatroid-optimization}
	Let \(\sigma\) and \(\rho\) satisfy the conditions of \thmref{intersection} and consider the LP given by
	\begin{equation}
		\label{eq:generalized-polymatroid}
		\begin{aligned}
			& \underset{R}{\text{minimize}} & & \sum_{s \in S}h(s)R(s)\\
			& \text{subject to} & & \sigma(U) \leq R(U) \leq \rho(U) & & U \subseteq S.
		\end{aligned}
	\end{equation}
	If \(h(s) \geq 0\) for all \(s \in S\), then there exists \(R^* \in \ext(B_\sigma)\) that is an optimal solution to the given LP.
	If \(h(s) \leq 0\) for all \(s \in S\), then there exists \(R^* \in \ext(B_\rho)\) that is an optimal solution to the given LP.
\end{cor}
For \(\sigma\) and \(\rho\) that satisfies the conditions of \thmref{intersection}, the set \(I_{\sigma, \rho}\) is a \emph{generalized polymatroid} \cite{FraTar1988}, a mathematical object that unifies polymatroids and contrapolymatroids \cite{Sch2003b}.
For every generalized polymatroid in \(\mathds{R}^{|S|}\), there exists a submodular set function \(\rho': 2^{|S| + 1} \rightarrow \mathds{R}\) and a projection \(p: \mathds{R}^{|S| + 1} \rightarrow \mathds{R}^{|S|}\) such that \(p(B_{\rho'})\) is equal to that generalized polymatroid \cite{Fuj2005} (see \fig{polymatroid-projection}).
This insight is half of the proof of \thmref{intersection}; the other half is recognizing that polyhedral properties are preserved by one-to-one affine mappings \cite{Fuj1984}.

We see from \eq{polymatroid-vertex} that the intersection \(I_{\sigma, \rho}\) has at most \((n+1)!\) vertices; we can construct trivial examples for which the intersection is a generalized polymatroid and has strictly less than \((n+1)!\) vertices.
For a given submodular set function \(\rho\), let \(\sigma(X) = \rho(S) - \rho(S \setminus X)\).
It can be readily verified that such a \(\sigma\) is supermodular and that \eq{intersection-thm} is always true by the submodularity of \(\rho\).
From \eq{generalized-polymatroid-vertices}, the vertices of \(I_{\sigma, \rho}\) are just those of \(B_\rho\) (or those of \(B_\sigma\) as \(B_\rho = B_\sigma\)).

We observe that when \eq{intersection-thm} holds, we have that \(B_\sigma \subset I_{\sigma, \rho}\) and \(B_\rho \subset I_{\sigma, \rho}\).
Motivated by the observation that if \((f^*, R^*)\) is an optimal solution to \eq{primal}, then \(R^* \in B_{\sigma_{SW}}\), we loosen the requirement \eq{intersection-thm} of \thmref{intersection} to characterize when \(B_\sigma \subset I_{\sigma, \rho}\).
\begin{thm}
	\label{thm:sigma-face}
	\(\ext(B_\sigma) \subseteq P_\rho\) if and only if
	\begin{equation}
		\label{eq:sigma-face}
		\sigma(T) - \sigma(T \setminus U) \leq \rho(U) \; \forall \, U \subseteq T \subseteq S.
	\end{equation}
\end{thm}
\begin{IEEEproof}
	\iftoggle{proofs}{\appendixproof{sigma-face-proof}}{Omitted for brevity.}
\end{IEEEproof}
Unsurprisingly, we can loosen  \eq{intersection-thm} in a similar manner to characterize when \(B_\rho \subset I_{\sigma, \rho}\).
\begin{thm}
	\label{thm:rho-face}
	\(\ext(B_\rho) \subseteq Q_\sigma\) if and only if
	\begin{equation}
		\label{eq:rho-face}
		\sigma(U) \leq \rho(T) - \rho(T \setminus U) \; \forall \,  U \subseteq T \subseteq S
	\end{equation}
\end{thm}
\begin{IEEEproof}
	\iftoggle{proofs}{\appendixproof{rho-face-proof}}{Omitted for brevity.}
\end{IEEEproof}

\begin{figure}
	\centering
	\subfloat[]{\includegraphics{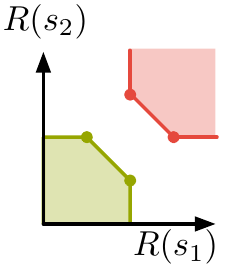}\label{fig:intersection-types-a}}
	\subfloat[]{\includegraphics{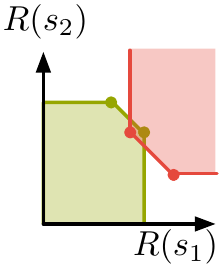}\label{fig:intersection-types-b}}
	\subfloat[]{\includegraphics{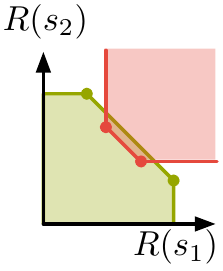}\label{fig:intersection-types-c}}\\
	\subfloat[]{\includegraphics{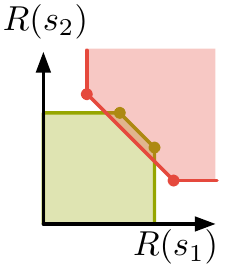}\label{fig:intersection-types-d}}
	\subfloat[]{\includegraphics{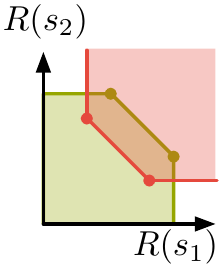}\label{fig:intersection-types-e}}
	\subfloat[]{\includegraphics{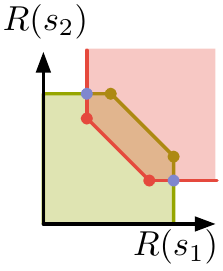}\label{fig:intersection-types-f}}\\
	\subfloat[]{\includegraphics[width=252.0pt]{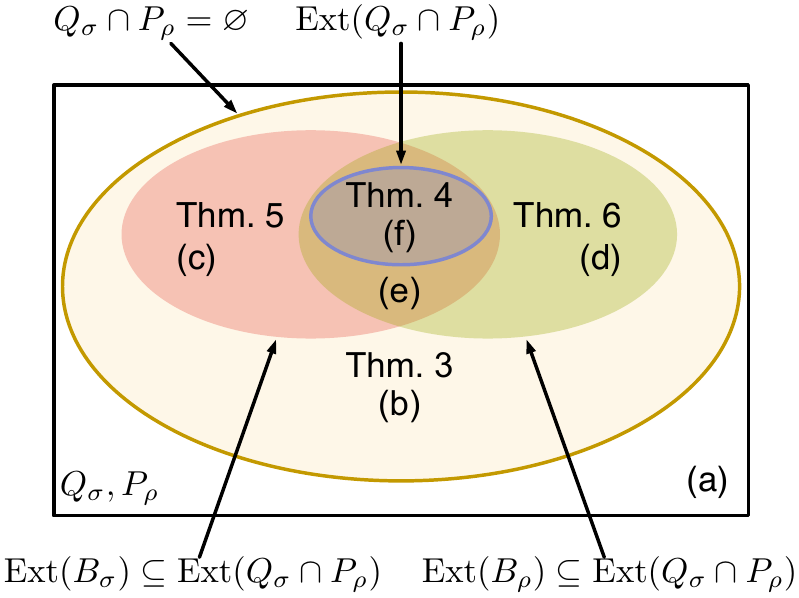}\label{fig:intersection-theorem-relationships}}
	\caption[]{%
		An example of Theorems \ref{thm:han-matching}, \ref{thm:intersection}, \ref{thm:sigma-face}, \& \ref{thm:rho-face}: %
		\thmref{han-matching} differentiates \subref{fig:intersection-types-a} vs.\ \subref{fig:intersection-types-b}--\subref{fig:intersection-types-e}. %
		If \thmref{han-matching} holds, \thmref{sigma-face} differentiates \subref{fig:intersection-types-b} vs.\ \subref{fig:intersection-types-c}; %
		\thmref{rho-face} differentiates \subref{fig:intersection-types-b} vs.\ \subref{fig:intersection-types-d}, and; %
		Theorems \ref{thm:sigma-face} \& \ref{thm:rho-face} \emph{together} differentiate \subref{fig:intersection-types-b} vs.\ \subref{fig:intersection-types-e}. %
		If \thmref{intersection} holds, we have a complete characterization of all the vertices of \(I_{\sigma, \rho}\); Theorems~\ref{thm:sigma-face} \& \ref{thm:rho-face} are not enough to characterize the vertices (purple, \subref{fig:intersection-types-f}) that are not in \(B_\sigma\) or \(B_\rho\).%
	}
	\label{fig:intersection-types}
\end{figure}

\begin{figure}
	\centering
	\includegraphics{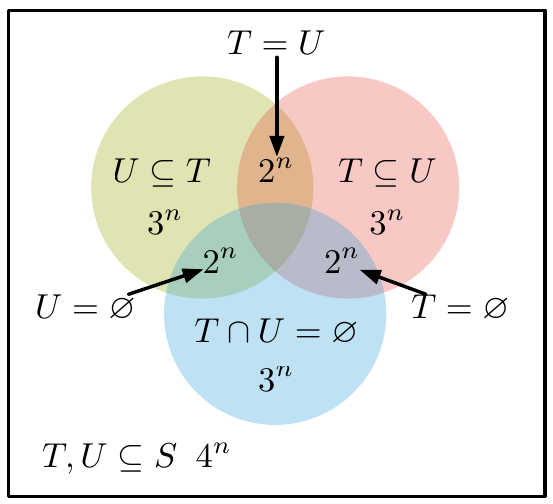}
	\caption{Relationship amongst Theorems \ref{thm:han-matching}, \ref{thm:intersection}, \ref{thm:sigma-face}, and \ref{thm:rho-face}: %
	Specializing the cross-inequality \eq{intersection-thm} of \thmref{intersection} for \(T = U\) gives \thmref{han-matching}, for \(T \subseteq U\) gives \thmref{sigma-face}, and for \(U \subseteq T\) gives \thmref{rho-face}.}
	\label{fig:cross-inequality}
\end{figure}

Observe that Theorems~\ref{thm:intersection}, \ref{thm:sigma-face}, and \ref{thm:rho-face} each imply \thmref{han-matching}.
To see this, let \(T = U\).
\fig{intersection-types} provides an example that illustrates the differences between Theorems~\ref{thm:han-matching}, \ref{thm:sigma-face}, \& \ref{thm:rho-face}.
\thmref{han-matching} provides an initial characterization of the structure of \(I_{\sigma, \rho}\) by determining when the intersection is empty or not and requires checking \(2^n\) inequalities.
It does not provide insight into what the vertices of the intersection are.
\thmref{sigma-face} and \thmref{rho-face} provide a partial characterization of the vertices of the intersection by characterizing a subset of the vertices of the intersection, but each requires checking \(3^n\) inequalities.
If \(\ext(B_\sigma) \subseteq P_\rho\), then \(\ext(B_\sigma) \subseteq \ext(I_{\sigma, \rho})\) and if \(\ext(B_\rho) \subseteq Q_\sigma\), then \(\ext(B_\rho) \subseteq \ext(I_{\sigma, \rho})\).
However, we know that there are vertices of \(I_{\sigma, \rho}\) that do not lie in either \(B_\sigma\) or \(B_\rho\) (e.g., \fig{intersection-types-e}).
Finally, \thmref{intersection} provides a complete characterization of \(\ext(I_{\sigma, \rho})\), but requires checking \(4^n\) inequalities.
Observe that \(U \cap T = \varnothing\) the cross inequality \eq{intersection-thm} of \thmref{intersection} is the tautology \(0 \leq 0\) and there are \(3^n\) such pairs \(T, U\) of subsets of \(S\).
For \(T = U\), \eq{intersection-thm} becomes Han's matching condition \eq{han-matching}.
For \(T \subseteq U\), \eq{intersection-thm} reduces to \eq{sigma-face}.
For \(U \subseteq T\), \eq{intersection-thm} reduces to \eq{rho-face}.
The relationship among Theorems~\ref{thm:intersection}, \ref{thm:sigma-face}, \& \ref{thm:rho-face} (in terms of pairs of subsets of \(S\)) is depicted in \fig{cross-inequality}.

\begin{figure}
	\centering
	\subfloat[]{\includegraphics{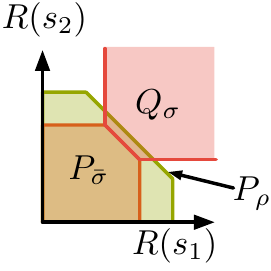}\label{fig:sigma-cross}}
	\subfloat[]{\includegraphics{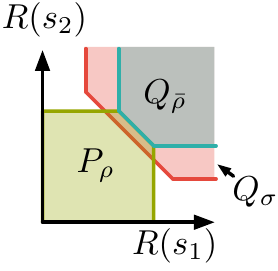}\label{fig:rho-cross}}
	\caption{Geometric interpretation of Theorems~\ref{thm:sigma-cross} and \ref{thm:rho-cross}. Since \(B_\sigma = B_{\bar{\sigma}}\), \(B_\sigma \subseteq P_\rho\) if and only if \(B_{\bar{\sigma}} \subseteq P_\rho\). A similar argument holds for \(B_\rho\).}
	\label{fig:simple-cross-example}
\end{figure}

We now show that characterizing \(B_\sigma \subset I_{\sigma, \rho}\) only requires checking \(2^n\) inequalities, as opposed to the \(3^n\) inequalities of \thmref{sigma-face}.
\begin{thm}
	\label{thm:sigma-cross}
	\begin{equation}
		\label{eq:sigma-cross-1}
		\sigma(T) - \sigma(T \setminus U) \leq \rho(U) \; \forall \, U \subseteq T \subseteq S
	\end{equation}
	if and only if
	\begin{equation}
		\label{eq:sigma-cross-2}
		\sigma(S) - \sigma(S \setminus U) \leq \rho(U) \; \forall \, U \subseteq S.
	\end{equation}

\end{thm}
\begin{IEEEproof}
	\iftoggle{proofs}{\appendixproof{sigma-cross-proof}}{Omitted for brevity.}
\end{IEEEproof}
While \eq{sigma-face} (and \eq{sigma-cross-1}) are readily seen as weaker versions of \eq{intersection-thm}, the relationship between \eq{sigma-cross-2} and \eq{intersection-thm} is not immediate.
As before, a similar result holds for characterizing \(B_\sigma \subset I_{\sigma, \rho}\).
\begin{thm}
	\label{thm:rho-cross}
	\begin{equation}
		\label{eq:rho-cross-1}
		\sigma(U) \leq \rho(T) - \rho(T \setminus U) \; \forall \, U \subseteq T \subseteq S
	\end{equation}
	if and only if
	\begin{equation}
		\label{eq:rho-cross-2}
		\sigma(U) \leq \rho(S) - \rho(S \setminus U) \; \forall \, U \subseteq S.
	\end{equation}
\end{thm}
\begin{IEEEproof}
	\iftoggle{proofs}{\appendixproof{rho-cross-proof}}{Omitted for brevity.}
\end{IEEEproof}
One advantage of Theorems~\ref{thm:sigma-cross} and \ref{thm:rho-cross} (besides the exponential reduction in inequalities), is the intuitive geometric interpretation of \eq{sigma-cross-2} and \eq{rho-cross-2}.
For a given supermodular set function \(\sigma\), let \(\bar{\sigma}(U) = \sigma(S) - \sigma(S \setminus)\), which is a submodular set function.
We then have (combining Theorems~\ref{thm:sigma-face} and \ref{thm:sigma-cross}) \(B_\sigma \subset I_{\sigma, \rho}\) if and only if \(\bar{\sigma}(U) \leq \rho(U)\) for all \(U \subset S\).
Equivalently the polymatroid \(P_{\bar{\sigma}}\) is a subset of the polymatroid \(P_\rho\), as depicted in \fig{simple-cross-example}.
With a similar argument combining Theorems~\ref{thm:rho-face} and \ref{thm:rho-cross} we have \(B_\sigma \subset I_{\sigma, \rho}\) if and only if the contrapolymatroid \(Q_{\bar{\rho}}\) is a subset of the contrapolymatroid \(Q_\sigma\).

We now specialize \eq{sigma-cross-2} for the case of conditional entropy \(\sigma_{SW}\) and min-cut capacity \(\rho_c\)
\begin{equation}
	H(X_U) \leq \rho_c(U) \quad \forall U \subseteq S,
\end{equation}
which follows from the application of the chain rule for entropy to \(H(X_S) - H(X_{S \setminus U} \mid X_U)\).
\begin{figure}
	\centering
	\includegraphics{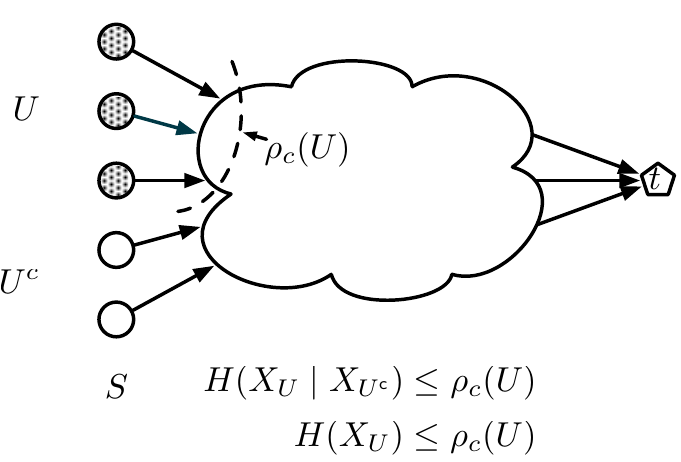}
	\caption{%
		Specializing \eq{han-matching} to conditional entropy and min-cut capacity requires the network to have enough capacity \(\rho_c(U)\) to support the lowest sum-rate \(H(X_U \mid X_{U^\mathsf{c}})\) from every subset of sources. %
		Specializing \eq{sigma-cross-2} to conditional entropy and min-cut capacity requires the network to have enough capacity \(\rho_c(U)\) to support the ``highest'' sum-rate \(H(X_U)\) from every subset of sources. %
		The sum-rate from a subset of sources could exceed the entropy, but is not needed to support lossless recovery of the sources and would be a suboptimal rate allocation.
	}
	\label{fig:sigma-cross-example}
\end{figure}
\fig{sigma-cross-example} illustrates the differences between \thmref{han-matching} and Theorems~\ref{thm:sigma-cross} and \ref{thm:rho-cross}.
Consider a set of sources \(U\).
Han's matching condition \eq{han-matching} requires the network have enough capacity to support the best-case sum-rate (i.e., minimum) from a set of sources for lossless recovery; in particular \(H(X_U  \mid X_{U^\mathsf{c}}) \leq \rho_c(U)\).
The matching condition \eq{sigma-cross-2} of \thmref{sigma-cross} requires that the network have enough capacity to support the worst-case sum-rate (i.e., maximum) for all subsets of sources.

Unlike \thmref{han-matching}, when we specialize \eq{intersection-thm} to the case of conditional entropy and min-cut capacity there is no immediately obvious intuition for what \eq{conditional-entropy-cross} represents.
\begin{equation}%
	\label{eq:conditional-entropy-cross}
	H(X_{U \cap T} \mid X_{U^\mathsf{c}}) \leq \rho_c(T) - \rho_c(T \setminus U) 
\end{equation}%

\subsection{Properties from conditional entropy}
\begin{figure}
	\centering
	\subfloat[]{\includegraphics{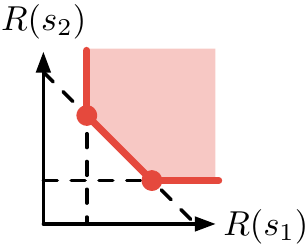}\label{fig:non-degenerate}}
	\subfloat[]{\includegraphics{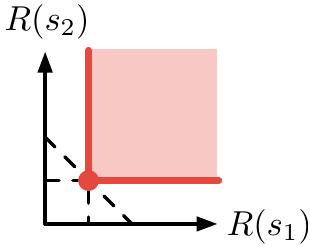}\label{fig:degenerate}}
	\caption[]%
	{At a non-degenerate vertex \subref{fig:non-degenerate}, \(|S|\) constraints will be active; a degenerate vertex \subref{fig:degenerate} will have more than \(|S|\) active constraints.}
	\label{fig:degeneracy}
\end{figure}

In the previous section, we focused on the properties of general submodular and supermodular set functions in order to more fully characterize the intersection of a polymatroid with a contrapolymatroid.
Our next set of results leverage additional properties of the conditional entropy supermodular set function, most notably the chain rule for entropy and the relationship between entropy and conditional independence \cite{CovTho2006}.
To motivate the results of this section, consider the two source SW rate regions in \fig{degeneracy}.
In general, the rate region is defined by three inequalities as in \fig{non-degenerate}; however, if it is known that the sources are independent (i.e., \(X_1 \perp X_2\)), the rate region can be defined using only two inequalities.
The number of vertices has also been reduced from two non-degenerate vertices to one \emph{degenerate} vertices.
We further develop this insight in the remainder of this section.

For an extreme point \(R\) of \(Q_{\sigma_{SW}}\), we provide an expression for the sum rate for an arbitrary set of sources and then use this to characterize the active inequalities of the LP \eq{primal} at \(R\).
\begin{lem}
	\label{lem:vertex-sum}
	Fix an ordering \(s_1, s_2, \ldots, s_n\) of the elements of \(S\) and define \(U_i \triangleq \left\{s_j : j \in [i]\right\}\).
	If \(R\) is the vertex in \(Q_{\sigma_{SW}}\) corresponding to this ordering and \(U = \{s_{k_1}, \cdots, s_{k_{m}}\}\) such that \(k_1 < k_2 < \ldots < k_m\) then
	\begin{equation}
		\begin{aligned}
			R(U) &= H(X_U | X_{U_{k_1}^\mathsf{c} \setminus U})\\
			     &+ \sum_{j = 2}^{m} I(X_{U \setminus U_{k_{j-1}}} ; X_{U_{k_j-1} \setminus U_{k_{j-1}}} | X_{U_{k_{j}}^\mathsf{c} \setminus U})
		\end{aligned}
	\end{equation}
\end{lem}
\begin{IEEEproof}
	\iftoggle{proofs}{\appendixproof{vertex-sum-proof}}{Omitted for brevity.}
\end{IEEEproof}
\begin{prop}
	\label{prop:active-constraint}
	Fix an ordering \(s_1, s_2, \ldots, s_n\) of the elements of \(S\) and define \(U_i \triangleq \left\{s_j : j \in [i]\right\}\) and \(U_0 = \varnothing\).
	Let \(R\) be the vertex in \(Q_{\sigma_{SW}}\) that corresponds to this ordering and \(U = \{s_{k_1}, \cdots, s_{k_{m}}\}\) such that \(k_1 < k_2 < \ldots < k_m\).
	Define \(k_0 \triangleq 0\).
	If \(U = U_i\) for some \(i \in [n]\), then \(R(U) = H(X_U | X_{U^\mathsf{c}})\).
	If \(U \neq U_i\) for some \(i\), then \(R(U) = H(X_U | X_U^\mathsf{c})\) if and only if
	\begin{equation}
		\label{eq:conditional-independence}
		(X_{U \setminus U_{k_{j-1}}} \perp X_{U_{k_{j}-1} \setminus U_{k_{j-1}}}) | X_{U_{k_{j}}^\mathsf{c} \setminus U} \quad j = 1, \ldots, m.
	\end{equation}
\end{prop}
\begin{IEEEproof}
	We provide a short summary here; see Appendix~\ref{sec:active-constraint-proof} for the proof in full detail.
	From the \lemref{vertex-sum}, we have that
	\begin{equation}
		\begin{aligned}
			0 &\leq R(U) - H(X_U | X_{U^\mathsf{c}})\\
			  &= \sum_{j = 1}^{m} I(X_{U \setminus U_{k_{j-1}}} ; X_{U_{k_j-1} \setminus U_{k_{j-1}}} | X_{U_{k_{j}}^\mathsf{c} \setminus U}).
		\end{aligned}
	\end{equation}
	The above is a sum of conditional mutual informations which is zero iff each of the terms is equal to zero.
	This happens when the random variables \(X_U\) satisfies \eq{conditional-independence}.
\end{IEEEproof}
\begin{prop}
	\label{prop:conditional-independence}
	Let \(T,U,V,W,Y\) be a partition of \(S\) and let \(\pi_T\) be a permutation of \(T\) etc.
	Define \(\pi = (\pi_T, \pi_U, \pi_V, \pi_W, \pi_Y)\) to be the permutation formed from the permutations of the associated partition and \(\pi' = (\pi_T, \pi_V, \pi_U, \pi_W, \pi_Y)\).
	If \(U \perp V \mid W\), then \(R_\pi = R_{\pi'}\).
\end{prop}
\begin{IEEEproof}
	\iftoggle{proofs}{\appendixproof{conditional-independence-proof}}{Omitted for brevity.}
\end{IEEEproof}
\begin{cor}
	Let \(T,U,V,W\) be a partition of \(S\) and let \(\pi_T\) be a permutation of \(T\) etc.
	Define \(\pi = (\pi_T, \pi_U, \pi_V, \pi_W)\) to be the permutation formed from the permutations of the associated partition and \(\pi' = (\pi_T, \pi_V, \pi_U, \pi_W)\).
	If \(U \perp V\), then \(R_\pi = R_{\pi'}\).
\end{cor}
A polyhedron can be represented as the intersection of half-spaces (\emph{H-rep}) or as the convex combination of its extreme points plus the conic combination of its extreme rays (\emph{V-rep})\cite{BerTsi1997}.
In general, it is more compact to represent polymatroids and contrapolymatroids using half-spaces (\(2^{|S|}\)) than in terms of the extreme points and extreme rays (\(\mathcal{O}(|S|!)\)).
What the previous two propositions show is that the size of the representation of the Slepian-Wolf rate region is directly tied to the conditional independence structure of the distributed correlated sources.
In turn, this means that the number of inequalities in the LP \eq{primal} depends on the conditional independence structure of the sources and may have a polynomial (in \(|S|\), \(|V|\), and \(|A|\)) number of constraints.
For example, if all the sources are independent, then only \(|S|\) inequalities are needed to describe the Slepian-Wolf rate region.

\begin{exmp}
	\label{exmp:markov-chain}
	Consider the following three source discrete memoryless source (DMS):
	\begin{equation}
		\mathds{P}(X_1 = x_1, X_2 = x_2) =
		\begin{dcases}
			\frac{1-p}{2} & x_1 = x_2\\
			\frac{p}{2} & x_1 \neq x_2
		\end{dcases}
	\end{equation}
	and
	\begin{equation}
		\mathds{P}(X_3 = 0 \mid X_2 = 0) = \mathds{P}(X_3 = 1 \mid X_2 = 1) = 1 - q
	\end{equation}
	with \(p, q \neq \frac{1}{2}\).
	Such a DMS forms the Markov chain \(X_1 \leftrightarrow X_2 \leftrightarrow X_3\).
	For the permutation \(\pi = (3, 1, 2)\), we have the three necessarily active constraints
	\begin{subequations}
		\begin{align}
			R_\pi(s_3) &= H(X_3 \mid X_1, X_2)\\
			R_\pi(\{s_1, s_3\}) &= H(X_1, X_3 \mid X_2)\\
			R_\pi(\{s_1, s_2, s_3\}) &= H(X_1, X_2, X_3).
		\end{align}
	\end{subequations}
	Additionally, because of the Markov structure for this source we have
	\begin{equation}
		R_\pi(s_1) = H(X_1 \mid X_2) = H(X_1 \mid X_2, X_3)
	\end{equation}
	active at \(R_\pi\).
	Enumerating all of the vertices, we see that the Slepian-Wolf rate region for this \emph{class} of DMSs has only five vertices instead of \(3! = 6\).
	In particular, the permutations \(\pi_1 = (1, 3, 2)\) and \(\pi_2 = (3, 1, 2)\) map to the same point, i.e., \(R_{\pi_1} = R_{\pi_2}\).
	
	If \(p = \frac{1}{2}\) or (exclusively) \(q = \frac{1}{2}\), then the SW rate region will only have four vertices.
	If \(p = q = \frac{1}{2}\), then the SW rate region will have one vertex.
	\begin{figure}
		\centering
		\includegraphics[width=252.0pt]{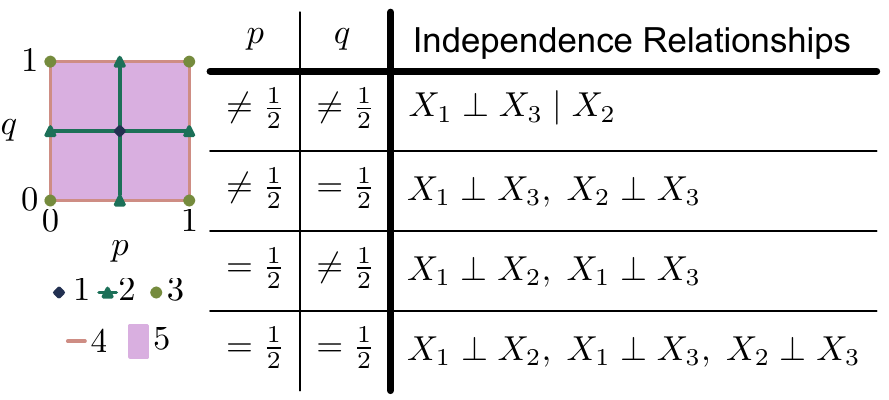}
		\caption{The number of vertices the Slepian-Wolf rate region has for the source of \exmpref{markov-chain} as a function of the parameters \(p\) and \(q\). Depending on the parameter \(p\) and \(q\), certain conditional independence relationships hold leading to a reduction in the number of vertices.}
		\label{fig:markov-chain-example}
	\end{figure}
	\fig{markov-chain-example} shows the number of vertices that SW rate region has and summarizes the different conditional independence structures as a function of the distribution parameters \(p\) and \(q\).
	\qed
\end{exmp}
\propref{active-constraint} will be used in the next section when giving conditions for a feasible solution of the optimization problem \eq{primal} to be optimal.

\section{Sufficient Conditions for Characterizing Optimality}
\label{sec:sufficient-conditions}

We proceed by finding the dual LP of the primal given in \eq{primal}.
In \eq{primal}, we have three types of constraints:
\begin{inparaenum}[i)]
\item a capacity constraint for each edge,
\item flow conservation for each node, and
\item rate requirements for each subset of sources.
\end{inparaenum}
The dual, then, will have three types of dual variables:
\begin{inparaenum}[i)]
\item \((x(a) : a \in A)\),
\item \((z(v) : v \in V)\), and
\item \((y_U : U \subset S)\).
\end{inparaenum}
The dual LP is given as
\begin{equation}
	\label{eq:dual}
	\begin{aligned}
		& \underset{x \leq 0, y \geq 0, z}{\text{maximize}} & & \sum_{a \in A}c(a)x(a) + \sum_{U \subseteq S}H(X_U|X_{U^\mathsf{c}})y_U & &\\
		& \text{subject to} & & x(a) + z(\head(a)) - z(\tail(a)) \leq k(a) & & a \in A\\
		& & & \sum_{U \ni s}y_U + z(s) - z(t) = h(s) & & s \in S
	\end{aligned}
\end{equation}
We set \(z(t) = 0\) because it is associated with the conservation of flow constraint at the sink, which is omitted from \eq{primal} as it is a consequence of the equality constraints at every other node.
Observe that the number of dual variables is exponential in \(|S|\).
We now show that, in a certain sense, the dual variables \(x(a)\) for \(a \in A\) and \(y_U\) for \(U \subseteq S\) are unnecessary.

Let us define the \emph{reduced cost} of \(a \in A\) as
\begin{equation}
	\label{eq:reduced-arc-cost}
	\bar{k}(a) \triangleq k(a) - (z(\head(a)) - z(\tail(a)))
\end{equation}
and observe that the first set of constraints of \eq{dual} can be written as \(x(a) \leq \bar{k}(a)\) for all \(a \in A\) \cite{CooCunPul2011}.
Combined with the non-positivity constraint on \(x(a)\) we have \(x(a) \leq \min(0, \bar{k}(a))\).
Since we are maximizing in \eq{dual} and \(c(a) > 0\) for all \(a\), we take
\begin{equation}
	\label{eq:arc-dual}
	x(a) = \min(0, \bar{k}(a))
\end{equation}
and see that the dual variable \(x(a)\) can be expressed in terms of \((z(v) : v \in V)\).
As we show in the next theorem, characterizations of optimal solutions do not need to explicitly consider the dual variables \((x(a) : a \in A)\).
\begin{thm}
	\label{thm:alternative-conj}
	Let \(f^*_{R_i}\) be a min-cost flow that supports rate \(R_i\). Let \(R = \sum_i \lambda_i R_i\).
	The flow \(f = \sum_i \lambda_i f^*_{R_i}\) is a flow that supports \(R\) of minimum cost if there exists a vector \((z(v) : v \in V)\) such that for all \(i\)
	\begin{subequations}
		\begin{align}
			\bar{k}(a) < 0 &\implies f^*_{R_i}(a) = c(a)\\
			\bar{k}(a) > 0 &\implies f^*_{R_i}(a) = 0.
		\end{align}
	\end{subequations}
\end{thm}
\begin{IEEEproof}
	\iftoggle{proofs}{\appendixproof{alternative-conj-proof}}{Omitted for brevity.}
\end{IEEEproof}
Since we are considering fixed rates in the previous theorem, there are no dual variables \((y_U : U \subseteq S)\).
If the conditional entropies of the sources and the min-cut capacities satisfy the requirements of \thmref{intersection}, then all extreme points of \(Q_{\sigma_{SW}}\) are feasible for \eq{primal}.
As was mentioned earlier, if \((f^*, R^*)\) is an optimal solution to \eq{primal} then \(R^*(S) = H(X_S)\) \cite{Han1980} and therefore \(R^*\) can be written as a convex combination of the extreme points of \(Q_{\sigma_{SW}}\).
The previous theorem shows that in certain cases, \(f^*\) can be found as a convex combination of the min-cost flows for the extreme points of the SW rate region.
In general though, this is not always the case as the next example demonstrates.
\begin{exmp}
	We consider the relay network with arc capacities and costs as shown in \fig{counter-example-solution-a}.
	Let the sources \(X_1, X_2\) be binary valued with the following joint distribution
	\begin{equation}
		\mathds{P}(X_1 = x_1, X_2 = x_2) =
		\begin{dcases}
			\frac{1-p}{2} & x_1 = x_2\\
			\frac{p}{2} & x_1 \neq x_2
		\end{dcases}.
	\end{equation}
	For such a source, the entropies are \(H(X_1) = H(X_2) = 1\) and \(H(X_1, X_2) = 1 + H(p)\) and the vertices of the Slepian-Wolf rate region are \(R_1 = (H(p), 1)\) and \(R_2 = (1, H(p))\)
	\footnote{\(H(p) = -p \log_2 p - (1 - p) \log_2(1 - p)\)}.
	The network of \fig{counter-example-solution-a} has sufficient capacity to support either \(R_1\) or \(R_2\).
	\begin{figure*}
		\centering
		\subfloat[]{\includegraphics{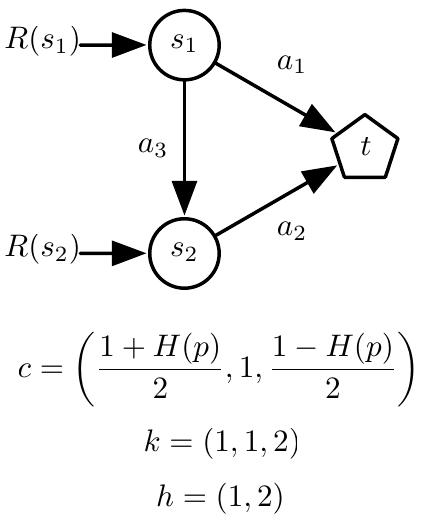}\label{fig:counter-example-solution-a}}
		\subfloat[]{\includegraphics{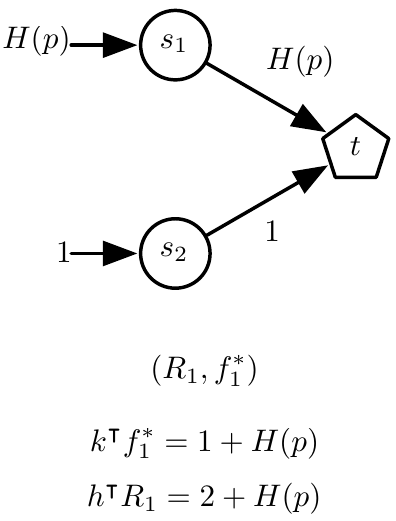}\label{fig:counter-example-solution-b}}
		\subfloat[]{\includegraphics{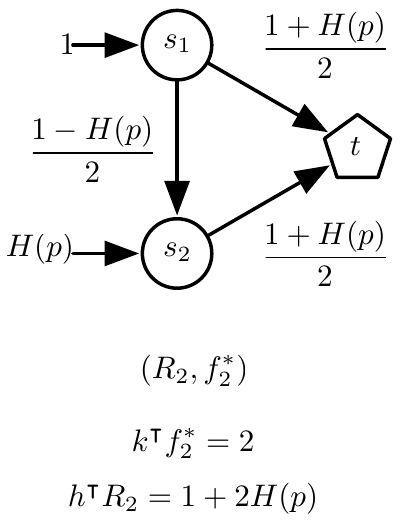}\label{fig:counter-example-solution-c}}\\
		\subfloat[]{\includegraphics{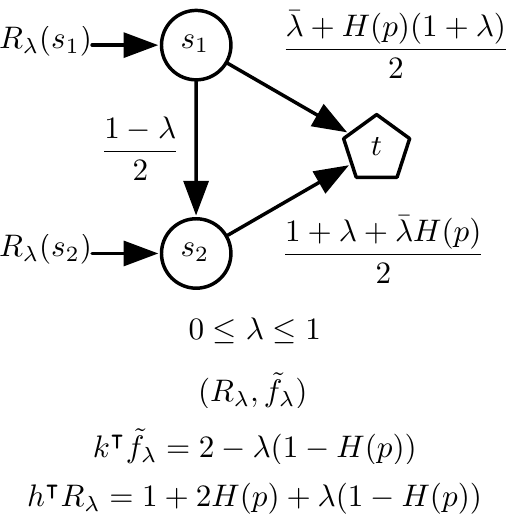}\label{fig:counter-example-solution-d}}
		\subfloat[]{\includegraphics{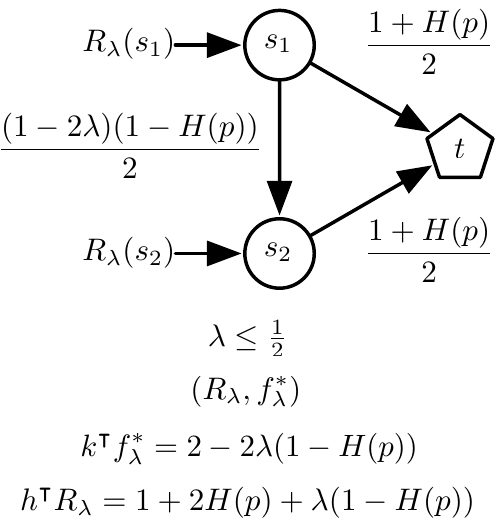}\label{fig:counter-example-solution-e}}
		\subfloat[]{\includegraphics{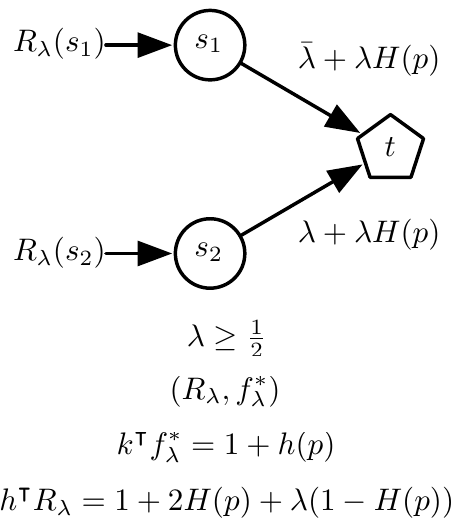}\label{fig:counter-example-solution-f}}
		\caption[]{
			Relay network example: 
			\subref{fig:counter-example-solution-a} network topology with arc capacities as a function of the source parameter \(p\), arc costs, and source costs; %
			\subref{fig:counter-example-solution-b} optimal min-cost flow when the source rates are fixed as \(R(s_1) = H(p)\) and \(R(s_2) = 1\); %
			\subref{fig:counter-example-solution-c} optimal min-cost flow when the source rates are fixed as \(R(s_1) = 1\) and \(R(s_2) = H(p)\); %
			\subref{fig:counter-example-solution-d} convex combination of \subref{fig:counter-example-solution-b} and \subref{fig:counter-example-solution-c}; %
			\subref{fig:counter-example-solution-e} optimal min-cost flow for convex combination of source rates in \subref{fig:counter-example-solution-b} and \subref{fig:counter-example-solution-c} for \(\lambda \leq \tfrac{1}{2}\), and; %
			\subref{fig:counter-example-solution-f} optimal min-cost flow for convex combination of source rates in \subref{fig:counter-example-solution-b} and \subref{fig:counter-example-solution-c} for \(\lambda \geq \tfrac{1}{2}\).%
		}\label{fig:counter-example-solutions}
	\end{figure*}
	Fixing \(R = R_1\) and solving for the min-cost flow \(f^*_1\), we obtain the solution shown in \fig{counter-example-solution-b}; correspondingly, if we fix \(R = R_2\) and solve for the min-cost flow \(f^*_2\), we obtain the solution shown in \fig{counter-example-solution-c}.
	The feasible solution \(\tilde{f}_\lambda = \lambda f^*_1 + (1 - \lambda)f_2^*\) for \(R_\lambda = \lambda R_1 + (1 - \lambda) R_2\) is shown in \fig{counter-example-solution-d}.
	Comparing with optimal min-cost flow \(f^*_\lambda\) for \(R_\lambda\) shown in \fig{counter-example-solution-e} \& \fig{counter-example-solution-f}, we see that for \(\lambda \in (0, 1)\), the convex combination of min-cost flows \(\tilde{f}_\lambda\) is \emph{not} a min-cost flow for \(R_\lambda\).
	Shown in \fig{counter-example-flow} is the cost \(k^\intercal \tilde{f}_\lambda\) of the convex combination of min-cost flows as a function of \(\lambda\) compared to the cost \(k^\intercal f^*_\lambda\) for the min-cost flow for a convex combination of rates \(R_\lambda\).
	Comparing \fig{counter-example-solution-d} with \fig{counter-example-solution-e} \& \fig{counter-example-solution-f}, we see immediately why \(\tilde{f}_\lambda\) is not optimal: \(\tilde{f}_\lambda\) always utilizes the arc \(a_3\) even when arc \(a_1\) (which has a lower cost) has spare capacity.
	If the same relay network is consider with \(k = (2, 1, 1)\) and all other parameters kept the same, then it can be shown that \(k^\intercal \tilde{f}_\lambda = k^\intercal f^*_\lambda\) for \(\lambda \in [0, 1]\).
	Observe that with this cost vector \(k\), the cost of the two directed paths \(s_1 \to t\) are \(k(a_1) = 2\) and \(k(a_3) + k(a_2) = 2\) and the cost of any flow supporting \(R(s_1)\) is the same.
	\begin{figure}
		\centering
		\includegraphics[width=252.0pt]{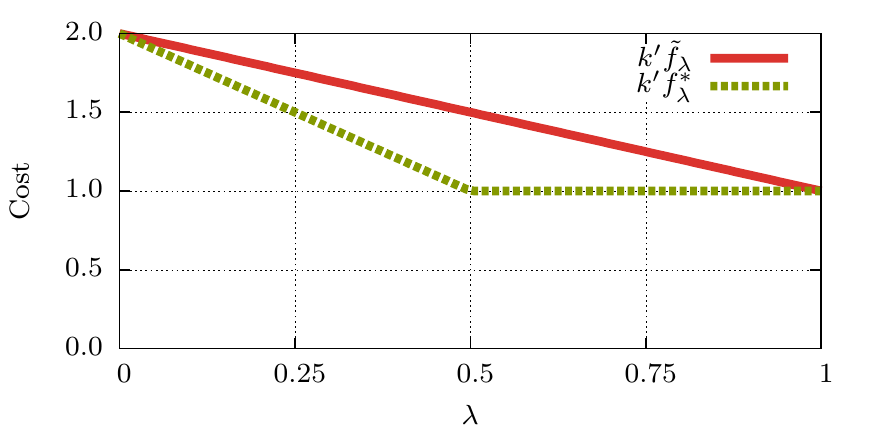}
		\caption{Plot of cost of a convex combination of min-cost flows and the cost of a min-cost flow for a convex combination of rates. The source distribution parameter \(p = \tfrac{1}{2}\).}\label{fig:counter-example-flow}
	\end{figure}
	\qed
\end{exmp}

A sufficient condition for the existence of \(z\) that satisfies the condition of \thmref{alternative-conj} can be given in terms of the topology of the network and the arc costs \(k\).
\begin{thm}
	\label{thm:equal-path-costs}
	If for every \(v \in V \setminus \{t\}\), the cost of all \(v-t\) paths are equal, then there exists a vector \((z(v) : v \in V)\) such that for
	\begin{subequations}
		\begin{align}
			\bar{k}(a) < 0 &\implies f^*_{R_i}(a) = c(a) \quad \forall i\\
			\bar{k}(a) > 0 &\implies f^*_{R_i}(a) = 0 \quad \forall i.
		\end{align}
	\end{subequations}
\end{thm}
\begin{IEEEproof}
	\iftoggle{proofs}{\appendixproof{equal-path-costs-proof}}{Omitted for brevity.}
\end{IEEEproof}

We now define the reduced cost of \(s \in S\) as
\begin{equation}
	\label{eq:source-reduced-cost}
	\bar{h}(s) \triangleq h(s) - (z(s) - z(t)) = h(s) - z(s)
\end{equation}
and rewrite the second set of constraints of \eq{dual} as
\begin{equation}
	\sum_{U \ni s}y_U = \bar{h}(s).
\end{equation}
We seek to express the dual variables \(y_U\) as a function of the dual variables \(z(s)\) as we did for the dual variables \(x(a)\).
The following theorem provides a characterization of which of the dual variables \(y_U\) must be zero as a function of the correlation structure of the source random variables.
\begin{thm}
	Suppose \(R^*\) is primal optimal and \(y^*\) is dual optimal and let \(U = \{s_{k_1}, \cdots, s_{k_{m}}\}\) such that \(k_1 < k_2 < \ldots < k_m\).
	If \(R^*\) is a vertex of \(Q_{\sigma_{SW}}\) and there exists \(j \in [m]\) such that
	\begin{equation}
		(X_{U \setminus U_{k_{j-1}}} \not\perp X_{U_{k_{j}-1} \setminus U_{k_{j-1}}}) | X_{U_{k_{j}}^\mathsf{c} \setminus U}
	\end{equation}
	then \(y^*_U = 0\).
\end{thm}
\begin{IEEEproof}
	Follows immediately from complimentary slackness and \propref{active-constraint}.
\end{IEEEproof}

This characterization suggests the following sufficient condition for an extreme point \(R_\pi\) of the SW rate region \(Q_{\sigma_{SW}}\) and its associated min-cost flow \(f^*_\pi\) to be a solution to the LP in \eq{primal}.
\begin{thm}
	\label{thm:optimal-check}
	A feasible solution \((f^*_\pi, R_\pi)\) of \eq{primal} is optimal if there exists vectors \((z(v) : v \in V)\) satisfying
	\begin{subequations}
		\begin{align}
			\bar{k}(a) < 0 &\implies f^*_\pi(a) = c(a) \quad \forall a \in A\\
			\bar{k}(a) > 0 &\implies f^*_\pi(a) = 0 \quad \forall a \in A
		\end{align}
	\end{subequations}
	and
	\begin{equation}
		\bar{h}(s_1) \geq \bar{h}(s_{2}) \geq \cdots \geq \bar{h}(s_n) \geq 0
	\end{equation}
	where the elements of \(S\) are ordered according the permutation \(\pi\).
\end{thm}
\begin{IEEEproof}
	\iftoggle{proofs}{\appendixproof{optimal-check-proof}}{Omitted for brevity.}
\end{IEEEproof}
The impact of the previous two theorems is that even though the dual has an exponential number of variables, we need only consider a linear (in \(|V|\)) number of them.
Given \((z^*(v): v \in V)\), we can compute \((x^*(a) : a \in A)\) according to \eq{arc-dual} and \((y^*_U : U \subseteq S)\) according to \eq{subset-dual}.
The extreme points of the SW rate region are significant because codes that satisfy \(R(S) = H(X_S)\) can be constructed from codes for these points via time sharing.
By adding in a per source cost to the previous example, we demonstrate that such a \(z\) need not always exist.
\begin{exmp}
	Looking at \fig{counter-example-flow}, one might conjecture that an optimal solution \((f^*, R^*)\) to problem in \eq{primal} would have the property \(R^* \in \ext(B_{\sigma_{SW}})\); i.e., that an optimal rate will always coincide with a vertex of the Slepian-Wolf rate region and that \(z\) that satisfies the condition of the previous theorem will always exist.
	It is certainly true that \(f^*, R^*\) will be a vertex of the polyhedron in flow-rate space, the network of \fig{counter-example-solutions} example demonstrates that for certain choices of source costs \(h\) and arc costs \(k\) the optimal rate \(R^*\) need \emph{not} be a vertex of the Slepian-Wolf rate region.
	\begin{figure}
		\centering
		\includegraphics{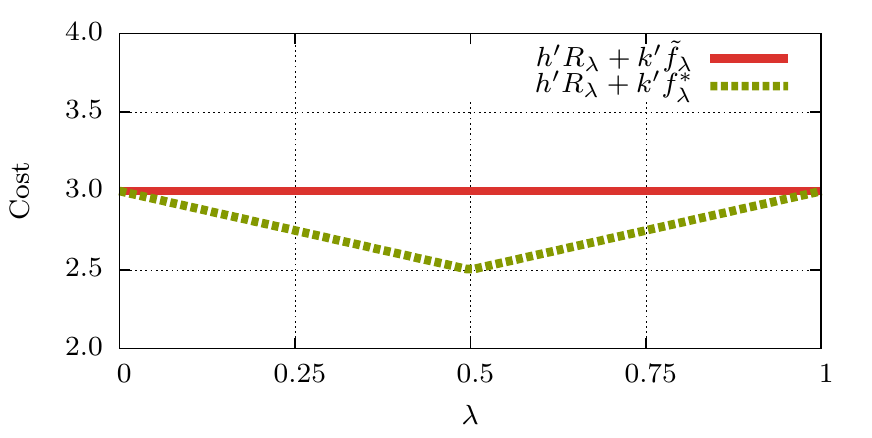}
		\caption{Plot of i) source cost plus cost of a convex combination of min-cost flows and ii) source cost plus the cost of a min-cost flow for a convex combination of rates. The source distribution parameter \(p = \tfrac{1}{2}\).}\label{fig:counter-example-cost}
	\end{figure}
	\fig{counter-example-cost} shows the cost \(h^\intercal R_\lambda + k^\intercal \tilde{f}_\lambda\) of the convex combination of min-cost flows as a function of \(\lambda\) compared to the cost \(h^\intercal R_\lambda + k^\intercal f^*_\lambda\) for the min-cost flow for a convex combination of rates \(R_\lambda\).
	We see immediately the minimum cost is achieved with \(\lambda = \tfrac{1}{2}\) and \(R^*_\lambda \not\in \ext(B_{\sigma_{SW}})\).
	\qed
\end{exmp}
Given the intuitive decomposition of the source coding and routing into different protocol layers noted by Barros et al., it may appear at first glance that a simple decomposed approach to designing a minimum cost solution might hold \cite{BarSer2006}.
For the case where all extreme points of the Slepian-Wolf rate region are feasible, one might consider a na\"{\i}ve approach of finding a minimum cost (w.r.t.\ \(h\)) source rate \(\hat{R}\) and a supporting minimum cost (w.r.t.\ \(k\)) \(\hat{f}\).
Alternatively, one might try enumerating all extreme points of the Slepian-Wolf rate region (combinatorial complexity aside), solving for a min-cost flow, and keeping track of the best solution.
The problem with both of these approaches is that the resulting feasible solution will select a rate \(\hat{R}\) that coincides with an extreme point of the Slepian-Wolf rate region.
The previous example demonstrates that when there is a imbalance between source costs and flow cost (i.e., cheap compression and expensive routing vs.\ expensive compression and cheap routing) the optimal rate \(R^* \not\in \ext(Q_{\sigma_{SW}})\).

\section{Extensions to Multiple Sinks}
\label{sec:multiple-sinks}
In previous sections, we have focused our attention on the single-sink problem.
In many contexts, it may be necessary to recover the source \(X_S\) at multiple sinks \(T\).
As mentioned earlier, this problem was considered by Ramamoorthy \cite{Ram2011}.
When there are multiple sinks, routing is no longer sufficient for conveyance of the sources to the sinks; instead network coding is necessary.
In the general network coding case, the single flow variable on each edges is replaced by virtual flows, one for each each edge and the traffic carried is represented with a physical flow \cite{LunRatMed2006}.
Finally, Ramamoorthy augments the original graph by adding in a super source \(s^*\) and connecting this vertex to each of the sources \(s \in S\) with an edge of zero cost and capacity given by the entropy of the source \(H(X_s)\).
With this augmentation, the multi-sink linear program can be written as:
\begin{equation}
	\label{eq:multi-sink-primal}
	\begin{aligned}
		& \underset{f, p, R}{\text{minimize}} \qquad \sum_{a \in A}k(a)p(a) & \\
		& \text{subject to} & & & &\\
		& 0 \leq f^{(t)}(a) \leq p(a) \leq c^*(a) & a \in A, t \in T \\
		& f^{(t)}(\delta^{in}(v)) - f^{(t)}(\delta^{out}(v)) = \Delta^{(t)}(v) & v \in V \\
		& x^{(t)}((s^*,s)) \geq R^{(t)}(s) & s \in S, t \in T \\
		& R^{(t)}(U) \geq H(X_U | X_{U^\mathsf{c}}) &U \subseteq S, t \in T.
	\end{aligned}
\end{equation}
where
\begin{equation}
	c^*(a) = \begin{dcases}
		c(a) & a \in A\\
		H(X_s) & a = (s^*, s)
	\end{dcases}
\end{equation}
and
\begin{equation}
	\Delta(v) = \begin{dcases}
		-H(X_S) & v = s^*\\
		H(X_S) & v = t\\
		0 & \text{otherwise}
	\end{dcases}
\end{equation}
We can view the multi-sink problem as being the intersection of multiple single-sink problems.
Looking at \eq{multi-sink-primal}, we can see that \(f^{(t)}\) is a valid flow that supports \(R^{(t)}\) for the sink \(t\).
We can define a min-cost capacity set function for each of the sinks
\begin{equation}
	\rho^{(t)}_c(U) = \min\{c(\delta^{out}(X)) : U \subseteq X, t \in V \setminus X\}
\end{equation}
and we then see that \(R^{(t)} \in Q_{\sigma_{SW}} \cap P_{\rho^{(t)}_c}\).

Although the multi-sink scenario requires network coding (as compared to distributed source coding and routing for the single-sink scenario), many of the insights developed for the single-sink case carry over.
The first is characterizing the feasibility of \eq{multi-sink-primal} \cite{SonYeu2001,Ram2011,Han2011}.
\begin{thm}[Han's Matching Condition for Multiple Sinks \cite{Han2011}]
	\label{thm:han-multi-sink-matching}
	The sources \(X_S\) are transmissible across the network to the sinks \(T\) if and only if
	\begin{equation}
		\label{eq:han-multi-sink-matching}
		\sigma_{SW}(U) \leq \min_{t \in T}\rho^{(t)}_c(U) \forall U \subseteq S.
	\end{equation}
\end{thm}
Comparing \eq{han-matching} and \eq{han-multi-sink-matching}, we see that \thmref{han-multi-sink-matching} implies \thmref{han-matching} for every sink \(t \in T\).
We can extend \thmref{sigma-face} in a similar manner.
\begin{thm}
	\label{thm:sigma-face-multi-sink}
	All of the vertices of the Slepian-Wolf rate region \(Q_{\sigma_{SW}}\) are feasible for the multi-sink problem \eq{multi-sink-primal} if and only if
	\begin{equation}
		\label{eq:sigma-face-multi-sink}
		H(X_U) \leq \min_{t \in T}\rho^{(t)}_c(U) \; \forall U \subseteq S.
	\end{equation}
\end{thm}
\begin{IEEEproof}
	\iftoggle{proofs}{\appendixproof{sigma-face-multi-sink-proof}}{Omitted for brevity.}
\end{IEEEproof}

We can bound the optimal value of the multi-sink min-cost flow problem in \eq{multi-sink-primal} in terms of the optimal values for a collection of single-sink min-cost flow problems.
\begin{thm}
	\label{thm:multi-sink-bounds}
	Let \((f^*, p^*, R^*)\) be an optimal solution to \eq{multi-sink-primal} and \((\hat{f}^{(t)}, \hat{R}^{(t)})\) be an optimal solution to the single sink problem for \(t \in T\).
	We have
	\begin{equation}
		\max_{t \in T}\sum_{a \in A} k(a)\hat{f}^{(t)}(a) \leq \sum_{a \in A}k(a)p^*(a)
	\end{equation}
	and
	\begin{equation}
		\sum_{a \in A}k(a)p^*(a) \leq \sum_{a \in A} k(a)\max_{t \in T}\hat{f}^{(t)}(a).
	\end{equation}
\end{thm}
\begin{IEEEproof}
	\iftoggle{proofs}{\appendixproof{multi-sink-bounds-proof}}{Omitted for brevity.}
\end{IEEEproof}
\begin{exmp}
	\label{exmp:ram2011}
	As an example of the above bounds, we consider a multi-sink problem instance formulated by Ramamoorthy \cite{Ram2011}.
	In this scenario, the network consists of \(50\) nodes and \(286\) edges; \(10\) of the nodes are sources and \(3\) are sinks, with the rest of the nodes acting as relays.
	The edge capacities are either \(20\) or \(40\) depending upon the distance between the connected nodes and the edge costs are all \(1\).
	A more detailed description of the network topology and source model can be found in \cite{Ram2011}, \S3-B.
	Shown in \fig{cost-iterations} is a comparison of the cost of the solution found using a partial dual decomposition and application of the subgradient method, the optimal value as computed from building and solving the LP in \eq{multi-sink-primal}, and the lower and upper bounds of \thmref{multi-sink-bounds}.\footnote{
		In Fig.\ 1 (b) of \cite{Ram2011}, the optimal value is reported as 646.69 which is different from the optimal value of 573.45 shown in \fig{cost-iterations}.
		Through personal correspondence with the author of \cite{Ram2011} it was determined that in computing the solution costs Fig.\ 1 (b) of \cite{Ram2011} non-zero costs were being assigned to the edges between the super source \(s^*\) and the sources \(s \in S\).
		Correcting for this gives the subgradient results in \fig{cost-iterations}
	}
	\begin{figure}
		\centering
		\includegraphics{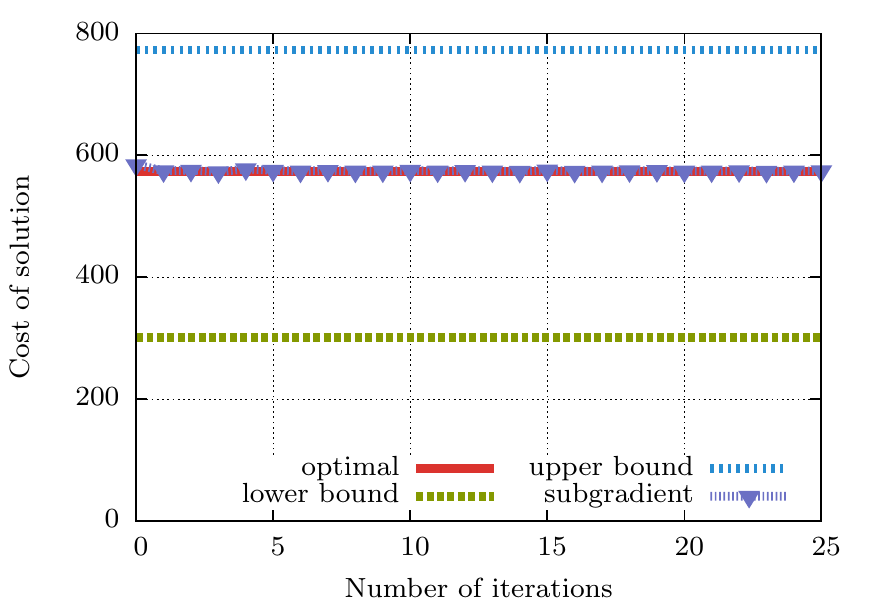}
		\caption{Comparison of subgradient method \cite{Ram2011} with optimal value and upper and lower bound.}
		\label{fig:cost-iterations}
	\end{figure}
	We observe, for this problem, that neither bound is tight; the lower bound has a relative difference of \(47.5\%\) while the upper bound has a relative difference of \(34.7\%\).
	We see that the subgradient method does converge to the optimal value rather quickly, realizing a relative error of 0.19\%, 0.13\%, and 0.06\% after 10, 100, and 1000 iterations, respectively.
	\qed
\end{exmp}

\section{Conclusion}
\label{sec:conclusion}
In this paper, we have considered the transmission of distributed sources across a network with capacity constraints.
Previous works have only made use of the fact that SW rate region is a contrapolymatroid as part of an iterative subgradient method.
The set of achievable rates is the intersection of the SW rate region with the polymatroid defined by the min-cut capacities.
We characterize when the SW vertices are all feasible and give an explicit characterization of all the vertices of the intersection of polymatroid with a contrapolymatroid for certain sub-/supermodular set functions.
The size of the representation of the SW rate region is related to the conditional independence relationships among the sources and in some cases may require a sub-exponential number of inequalities to describe the rate region.
We have shown that these properties lead to a characterization relating optimal solutions and the corner points of the SW rate region.
Through a simple, but natural counter-example we demonstrate that an optimal rate allocation may not be a vertex of the SW rate region.
Our result concerning the feasibility of all the SW rate region vertices naturally extends from the single sink problem to the multi-sink setting.
The optimal value of the multi-sink is bounded from above and below in terms of the optimal solutions to a collection of related single-sink problems.

\bibliographystyle{IEEEtran}
\bibliography{sources}

\section*{Acknowledgment}
The authors would like to thank A. Ramamoorthy for graciously providing data utilized in producing \fig{cost-iterations}.

\section*{Disclaimer}
The views and conclusions contained herein are those of the authors and should not be interpreted as necessarily representing the official policies or endorsements, either expressed or implied, of the Air Force Research Laboratory or the U.S.\ Government.

\iftoggle{proofs}{\appendix
\allowdisplaybreaks
\subsection{Included Proofs}
\subsubsection{Proof of \lemref{vertex-sum-chain}}
\label{sec:vertex-sum-chain-proof}
\begin{IEEEproof}
	For any supermodular set function we have
	\begin{equation}
		\begin{aligned}
			R_\pi(U_{\pi(i)}) &= \sum_{j = 1}^i R_\pi(s_{\pi(j)})\\
					  &= \sum_{j = 1}^i \sigma(U_{\pi(j)}) - \sigma(U_{\pi(j-1)})\\
					  &= \sigma(U_{\pi(i)}) - \sigma(\varnothing).
		\end{aligned}
	\end{equation}
	For any submodular set function we have
	\begin{equation}
		\begin{aligned}
			R_\pi(U_{\pi(i)}) &= \sum_{j = 1}^i R_\pi(s_{\pi(j)})\\
					  &= \sum_{j = 1}^i \rho(U_{\pi(j)}) - \rho(U_{\pi(j-1)})\\
					  &= \rho(U_{\pi(i)}) - \rho(\varnothing).
		\end{aligned}
	\end{equation}
\end{IEEEproof}

\subsubsection{Proof of \thmref{intersection}}
\label{sec:intersection-proof}
\begin{lem}
	\label{lem:projection}
	Let \(\ext(P)\) be the set of extreme points of a polyhedron \(P\) and \(p(P)\) be the projection of \(P\); then \(\ext(p(P)) \subseteq p(\ext(P))\).
\end{lem}
\begin{IEEEproof}
	Suppose \(x \in \ext(p(P))\); then there exists \(x'\) such that \((x, x') \in P\).
	As \(x\) is extreme, there does not exist \(y, z \in p(P)\) not equal to \(x\) and \(\lambda \in (0, 1)\) such that \(x = \lambda y + (1 - \lambda)z\).
	Therefore there is no choice of \((y, y'), (z, z') \in P\) not equal to \((x, x')\) and \(\lambda \in (0, 1)\) such that
	\begin{equation}
		\begin{pmatrix} x \\ x' \end{pmatrix}
		= \lambda \begin{pmatrix} y \\ y' \end{pmatrix}
		+ (1 - \lambda)	\begin{pmatrix} z \\ z' \end{pmatrix}
	\end{equation}
	and \((x, x') \in \ext(P)\).
\end{IEEEproof}
\begin{lem}
	\label{lem:affine-projection}
	Let \(\ext(P)\) be the set of extreme points of a polyhedron \(P\) and \(p(P)\) be the projection of \(P\).
	If \(p\) one-to-one, then \(\ext(p(P)) = p(\ext(P))\).
\end{lem}
\begin{IEEEproof}
	Consider \(x \in \ext(P)\); suppose its projection \(p(x) \not\in \ext(p(P))\).
	W.l.o.g.\ there exists \(y, z \in P\) not equal to \(x\) and \(\lambda \in (0, 1)\) such that
	\begin{equation}
		p(x) = \lambda p(y) + (1 - \lambda) p(z) = p(\lambda y + (1 - \lambda) z)
	\end{equation}
	which follows from projections being affine mappings.
	Additionally, since the projection is one-to-one we must have
	\begin{equation}
		x = \lambda y + (1 - \lambda) z
	\end{equation}
	contradicting the assumption of \(x \in \ext(P)\).
\end{IEEEproof}
\begin{IEEEproof}
	Assuming \(\sigma\) and \(\rho\) satisfy the condition of \thmref{intersection}, we have that \(Q_{\sigma} \cap P_{\rho}\) is non-empty.
	Let us define \(S' = S \cup \{s^*\}\) and
	\begin{equation}
		f(U) = \begin{dcases}
			\rho(U) & U \in 2^S\\
			\gamma - \sigma(S' \setminus U) & U \subset S', s^* \in U
		\end{dcases}
	\end{equation}
	where \(\gamma \in \mathds{R}\) is arbitrary but fixed.
	Such a \(f\) is a submodular function on \(2^{S'}\) and \(B(EP_f) = EP_f \cap \{R : R(S) = f(S)\}\)
	\footnote{The set \(EP_f = \{R \in \mathds{R}^{|S'|} : R(U) \leq f(U)\}\) is the extended polymatroid associated with \(f\) while \(P_f = \{x \in \mathds{R}^{|S'|}: R \geq 0, R(U) \leq f(U)\}\) is the polymatroid associated with \(f\).}
	is non-empty \cite{Fuj2005}.
	In fact
	\begin{equation}
		Q_{\sigma} \cap P_{\rho} = \left\{R \in \mathds{R}^{|S|} : \exists \alpha \in \mathds{R} : (R, \alpha) \in B(EP_f)\right\}.
	\end{equation}
	The vertices of \(EP_f\) are given by
	\begin{equation}
		R_\pi(s_{\pi(i)}) = f(U_{\pi(i)}) - f(U_{\pi(i-1)}) \quad i = 1, \ldots, |S| + 1
	\end{equation}
	where \(\pi\) ranges over all permutations of \([|S| + 1]\).
	Let \(j\) be the integer such that \(s_{\pi(j)} = s'\).
	Then
	\begin{equation}
		R_\pi(s_{\pi(i)}) =
		\begin{dcases}
			\rho(U_{\pi(i)}) - \rho(U_{\pi(i-1)}) & i < j\\
			\gamma - \sigma(S' \setminus U_{\pi(i)}) - \rho(U_{\pi(i-1)}) & i = j\\
			\sigma(S' \setminus U_{\pi(i-1)}) - \sigma(S' \setminus U_{\pi(i)}) & i > j.
		\end{dcases}
	\end{equation}
\end{IEEEproof}

\subsubsection{Proof of \thmref{sigma-face}}
\label{sec:sigma-face-proof}
\begin{IEEEproof}
	Assume \(\sigma(T) - \sigma(T \setminus U) \leq \rho(U)\) for all \(U \subseteq T \subseteq S\).
	Consider an arbitrary permutation \(\pi\) and its associated vertex \(R_\pi\) of \(Q_\sigma\).
	For any \(U \subseteq S\), define \(k \triangleq \min \{k' : U \subseteq U_{\pi(k')}\}\) or equivalently \(k \triangleq \max\{k' : s_{\pi(k')} \in U\}\).
	We have
	\begin{equation}
	\begin{aligned}
		R_\pi(U) &= R_\pi(U_{\pi(k)}) - R_\pi(U_{\pi(k)} \setminus U)\\
			 &= \sigma(U_{\pi(k)}) - R_\pi(U_{\pi(k)} \setminus U)\\
			 &\leq \sigma(U_{\pi(k)}) - \sigma(U_{\pi(k)} \setminus U)\\
			 &\leq \rho(U)
	\end{aligned}
	\end{equation}
	and therefore \(R_\pi \in P_\rho\).
	This is true for all permutations and we conclude that \(B(Q_\sigma) \subseteq P_\rho\).

	Suppose \(\exists U \subseteq T \subseteq S\) such that \(\sigma(T) - \sigma(T \setminus U) > \rho(U)\).
	Let the elements of \(S\) be ordered by a permutation \(\pi\) so that \(T = \{s_{\pi(1)}, \ldots, s_{\pi(|T|)}\}\) and \(U = \{s_{\pi(|T|-|U|+1)}, \ldots, s_{\pi(|T|)}\}\).
	Then \(T = U_{\pi(|T|)}\) and \(T \setminus U = U_{\pi(|T|-|U|)}\).
	It follows that
	\begin{equation}
	\begin{aligned}
		R_\pi(U) &= \sum_{i = |T| - |U| + 1}^{|T|}\sigma(U_{\pi(i)}) - \sigma(U_{\pi(i-1)})\\
			 &= \sigma(U_{\pi(|T|)}) - \sigma(U_{\pi(|T|-|U|)})\\
			 &= \sigma(T) - \sigma(T \setminus U)\\
			 &> \rho(U)
	\end{aligned}
	\end{equation}
	and therefore \(R_\pi \not\in P_\rho\).
	We conclude that \(B_\sigma \not\subseteq P_\rho\).
\end{IEEEproof}

\subsubsection{Proof of \thmref{rho-face}}
\label{sec:rho-face-proof}
\begin{IEEEproof}
	Assume \(\sigma(U) \leq \rho(T) - \rho(T \setminus U\) for all \(U \subseteq T \subseteq S\).
	Consider an arbitrary permutation \(\pi\) and its associated vertex \(R_\pi\) of \(P_\rho\).
	For any \(U \subseteq S\), define \(k \triangleq \min \{k' : U \subseteq U_{\pi(k')}\}\) or equivalently \(k \triangleq \max\{k' : s_{\pi(k')} \in U\}\).
	We have
	\begin{equation}
	\begin{aligned}
		R_\pi(U) &= R_\pi(U_{\pi(k)}) - R_\pi(U_{\pi(k)} \setminus U)\\
			 &= \rho(U_{\pi(k)}) - R_\pi(U_{\pi(k)} \setminus U)\\
			 &\geq \rho(U_{\pi(k)}) - \rho(U_{\pi(k)} \setminus U)\\
			 &\geq \sigma(U)
	\end{aligned}
	\end{equation}
	and therefore \(R_\pi \in Q_\sigma\).
	This is true for all permutations and we conclude that \(B(P_\rho) \subseteq Q_\sigma\).

	Suppose \(\exists U \subseteq T \subseteq S\) such that \(\sigma(U) > \rho(T) - \rho(T \setminus U)\).
	Let the elements of \(S\) be ordered by a permutation \(\pi\) so that \(T = \{s_{\pi(1)}, \ldots, s_{\pi(|T|)}\}\) and \(U = \{s_{\pi(|T|-|U|+1)}, \ldots, s_{\pi(|T|)}\}\).
	Then \(T = U_{\pi(|T|)}\) and \(T \setminus U = U_{\pi(|T|-|U|)}\).
	It follows that
	\begin{equation}
	\begin{aligned}
		R_\pi(U) &= \sum_{i = |T| - |U| + 1}^{|T|}\rho(U_{\pi(i)}) - \rho(U_{\pi(i-1)})\\
			 &= \rho(U_{\pi(|T|)}) - \rho(U_{\pi(|T|-|U|)})\\
			 &= \rho(T) - \rho(T \setminus U)\\
			 &< \sigma(U)
	\end{aligned}
	\end{equation}
	and therefore \(R_\pi \not\in Q_\sigma\).
	We conclude that \(B_\rho \not\subseteq Q_\sigma\).
\end{IEEEproof}
\begin{rem}
	In the proofs of Theorems~\ref{thm:sigma-face} \& \ref{thm:rho-face}, we use the existence of \(T, U\) that do not satisfy \eq{sigma-face} (resp. \eq{rho-face}) to construct a vertex of \(Q_\sigma\) (resp. \(P_\rho\)) that is not retained in the intersection \(I_{\sigma, \rho}\).
	For a given \(T, U\) that do not satisfy \eq{sigma-face}, there exists \((|T| - |U|)!\,|U|!\,(|S| - |T|)!\) permutations for which the corresponding vertex of \(Q_\sigma\) is not in \(P_\rho\).
	Similarly, for a given \(T, U\) that do not satisfy \eq{rho-face}, there exists \((|T| - |U|)!\,|U|!\,(|S| - |T|)!\) permutations for which the corresponding vertex of \(P_\rho\) is not in \(Q_\sigma\).
\end{rem}

\subsubsection{Proof of \thmref{sigma-cross}}
\label{sec:sigma-cross-proof}
\begin{IEEEproof}
	The set of inequalities in \eq{sigma-cross-1} include \eq{sigma-cross-2}, so the one direction is immediate.

	By the supermodularity of \(\sigma\), we have
	\begin{equation}
		\sigma(T) + \sigma(S \setminus U) \leq \sigma(S) + \sigma(T \setminus U) \; \forall U \subseteq T \subset S
	\end{equation}
	which we rearrange to get
	\begin{equation}
		\sigma(T) - \sigma(T \setminus U) \leq \sigma(S) - \sigma(S \setminus U) \leq \rho(U) \; \forall U \subseteq T \subset S.
	\end{equation}
\end{IEEEproof}

\subsubsection{Proof of \thmref{rho-cross}}
\label{sec:rho-cross-proof}
\begin{IEEEproof}
	The set of inequalities in \eq{rho-cross-1} include \eq{rho-cross-2}, so the one direction is immediate.

	By the submodularity of \(\rho\), we have
	\begin{equation}
		\rho(S) + \rho(T \setminus U) \leq \rho(T) + \rho(S \setminus U) \; \forall U \subseteq T \subset S
	\end{equation}
	which we rearrange to get
	\begin{equation}
		\sigma(U) \leq \rho(S) - \sigma(S \setminus U) \leq \rho(T) - \rho(T \setminus U) \; \forall U \subseteq T \subset S.
	\end{equation}
\end{IEEEproof}

\subsubsection{Proof of \lemref{vertex-sum}}
\label{sec:vertex-sum-proof}
Recall from \lemref{vertex-sum} that \(U_i \triangleq \left\{s_j : j \in [i]\right\}\) and \(U = \{s_{k_1}, \cdots, s_{k_{m}}\}\) such that \(k_1 < k_2 < \ldots < k_m\).
Let us define \(U' \triangleq U \setminus \{s_{k_1}\} = \{s_{k'_1}, \cdots, s_{k'_{m'}}\}\) where \(k'_{i} = k_{i+1}\) and \(m' = m - 1\).
We begin with three supporting lemmas.
\begin{lem}
	\begin{equation}
		U_{k_j}^\mathsf{c} \setminus U' = U_{k_j}^\mathsf{c} \setminus U
	\end{equation}
\end{lem}
\begin{IEEEproof}
	\begin{equation}
	\begin{aligned}
		U_{k_j}^\mathsf{c} \setminus U &= U_{k_j}^\mathsf{c} \cap (\{s_{k_1}\} \cup U')^\mathsf{c}\\
					       &= U_{k_j}^\mathsf{c} \cap (\{s_{k_1}\}^\mathsf{c} \cap U^{'\mathsf{c}})\\
					       &= U_{k_j}^\mathsf{c} \cap U^{'\mathsf{c}}
	\end{aligned}
	\end{equation}
	The first step follows from the definition of \(U'\) and the last step from recognizing that \(U_{k_j}^\mathsf{c} \subseteq \{s_{k_1}\}^\mathsf{c}\).
\end{IEEEproof}
\begin{lem}
	\begin{equation}
		U' = U \setminus U_{k_1}
	\end{equation}
\end{lem}
\begin{IEEEproof}
	\begin{equation}
		U \setminus U_{k_1} = U \cap \{s_{k_1+1}, s_{k_1+2}, \ldots, s_n\} = \{s_{k_2}, \cdots, s_{k_{m}}\} = U'
	\end{equation}
\end{IEEEproof}
\begin{lem}
	\begin{equation}
		U^\mathsf{c} = U_{k_1-1} \cup U_{k_1}^\mathsf{c} \setminus U
	\end{equation}
\end{lem}
\begin{IEEEproof}
	\begin{equation}
		\begin{aligned}
			U_{k_1-1} \cup U_{k_1}^\mathsf{c} \setminus U &= U_{k_1-1} \cup (U_{k_1}^\mathsf{c} \cap U^\mathsf{c})\\
							     &= (U_{k_1-1} \cup U_{k_1}^\mathsf{c}) \cap (U_{k_1-1} \cup U^\mathsf{c})\\
							     &= \{s_{k_1-1}\}^\mathsf{c} \cap U^\mathsf{c}\\
							     &= U^\mathsf{c}
		\end{aligned}
	\end{equation}
\end{IEEEproof}

\begin{IEEEproof}[Proof of \lemref{vertex-sum}]
	Proof by induction on \(|U|\).
	\emph{Base case:} If \(|U| = 1\), then \(U = \{s_{k_1}\}\) and we have that
	\begin{equation}
		\begin{aligned}
			R(s_{k_1}) &= H(X_{U_{k_1}} \mid X_{U_{k_1}^\mathsf{c}}) - H(X_{U_{{k_1}-1}} \mid X_{U_{{k_1}-1}^\mathsf{c}})\\
				   &= H(X_{U_{{k_1}-1}}, X_{s_{k_1}} \mid X_{U_{k_1}^\mathsf{c}}) - H(X_{U_{{k_1}-1}} \mid X_{U_{{k_1}}^\mathsf{c}}, X_{s_{k_1}})\\
				   &= H(X_{s_{k_1}} \mid X_{U_{k_1}^\mathsf{c}})\\
				   &= H(X_{s_{k_1}} \mid X_{U_{k_1}^\mathsf{c} \setminus \{s_{k_1}\}})\\
		\end{aligned}
	\end{equation}
	where the last step follows from the fact that \(U_{i}^\mathsf{c} = U_{i}^\mathsf{c} \setminus \{s_{i}\}\).

	\emph{Inductive step:} Let us define
	\begin{equation}
		U' \triangleq U \setminus \{s_{k_1}\} = \{s_{k'_1}, \cdots, s_{k'_{m'}}\}
	\end{equation}
	where \(k'_{i} = k_{i+1}\) and \(m' = m - 1\).
	We have that
	\begin{align*}
		R(U) &= R(s_{k_1}) + R(U')\\
		     &\stackrel{(a)}{=} H(X_{s_{k_1}} | X_{U_{k_1}^\mathsf{c}}) + R(U')\\
		     &\stackrel{(b)}= H(X_{s_{k_1}} | X_{U_{k_1}^\mathsf{c}}) + H(X_{U'} | X_{U_{k'_1}^\mathsf{c} \setminus U'})\\
		     &+ \sum_{i = 1}^{m' - 1} I(X_{U' \setminus U_{k'_i}} ; X_{U_{k'_{i+1}-1} \setminus U_{k'_i}} | X_{U_{k'_{i+1}}^\mathsf{c} \setminus U'})\\
		     &\stackrel{(c)}{=} H(X_{s_{k_1}} | X_{U_{k_1}^\mathsf{c}})\\
		     &+ H(X_{U'} | X_{U_{k_2-1} \setminus U_{k_1}}, X_{U_{k'_1}^\mathsf{c} \setminus U'})\\
		     &+ I(X_{U'};X_{U_{k_2-1} \setminus U_{k_1}} | X_{U_{k'_1}^\mathsf{c} \setminus U'})\\
		     &+ \sum_{i = 1}^{m' - 1} I(X_{U' \setminus U_{k'_i}} ; X_{U_{k'_{i+1}-1} \setminus U_{k'_i}} | X_{U_{k'_{i+1}}^\mathsf{c} \setminus U'})\\
		     &\stackrel{(d)}{=} H(X_{s_{k_1}} | X_{U_{k_1}^\mathsf{c}})\\
		     &+ H(X_{U'} | X_{U_{k_2-1} \setminus U_{k_1}}, X_{U_{k_2}^\mathsf{c} \setminus U'})\\
		     &+ I(X_{U'};X_{U_{k_2-1} \setminus U_{k_1}} | X_{U_{k_2}^\mathsf{c} \setminus U'})\\
		     &+ \sum_{i = 1}^{m' - 1} I(X_{U' \setminus U_{k'_i}} ; X_{U_{k'_{i+1}-1} \setminus U_{k'_i}} | X_{U_{k'_{i+1}}^\mathsf{c} \setminus U'})\\
		     &\stackrel{(e)}{=} H(X_{s_{k_1}} | X_{U_{k_1}^\mathsf{c} \setminus U'}, X_{U'}) + H(X_{U'} | X_{U_{k_1}^\mathsf{c} \setminus U'})\\
		     &+ I(X_{U'};X_{U_{k_2-1} \setminus U_{k_1}} | X_{U_{k_2}^\mathsf{c} \setminus U'})\\
		     &+ \sum_{i = 1}^{m' - 1} I(X_{U' \setminus U_{k'_i}} ; X_{U_{k'_{i+1}-1} \setminus U_{k'_i}} | X_{U_{k'_{i+1}}^\mathsf{c} \setminus U'})\\
		     &\stackrel{(f)}{=} H(X_{U} | X_{U_{k_1}^\mathsf{c} \setminus U'})\\
		     &+ I(X_{U'};X_{U_{k_2-1} \setminus U_{k_1}} | X_{U_{k_2}^\mathsf{c} \setminus U'})\\
		     &+ \sum_{i = 1}^{m' - 1} I(X_{U' \setminus U_{k'_i}} ; X_{U_{k'_{i+1}-1} \setminus U_{k'_i}} | X_{U_{k'_{i+1}}^\mathsf{c} \setminus U'})\\
		     &\stackrel{(g)}{=} H(X_{U} | X_{U_{k_1}^\mathsf{c} \setminus U'})\\
		     &+ I(X_{U'};X_{U_{k_2-1} \setminus U_{k_1}} | X_{U_{k_2}^\mathsf{c} \setminus U'})\\
		     &+ \sum_{i = 2}^{m - 1} I(X_{U \setminus U_{k_i}} ; X_{U_{k_{i+1}-1} \setminus U_{k_i}} | X_{U_{k_{i+1}}^\mathsf{c} \setminus U})\\
		     &\stackrel{(h)}{=} H(X_{U} | X_{U_{k_1}^\mathsf{c} \setminus U'})\\
		     &+ I(X_{U \setminus U_{k_1}};X_{U_{k_2-1} \setminus U_{k_1}} | X_{U_{k_2}^\mathsf{c} \setminus U})\\
		     &+ \sum_{i = 2}^{m - 1} I(X_{U \setminus U_{k_i}} ; X_{U_{k_{i+1}-1} \setminus U_{k_i}} | X_{U_{k_{i+1}}^\mathsf{c} \setminus U})\\
		     &\stackrel{(i)}{=} H(X_{U} | X_{U_{k_1}^\mathsf{c} \setminus U})\\
		     &+ \sum_{i = 1}^{m - 1} I(X_{U \setminus U_{k_i}} ; X_{U_{k_{i+1}-1} \setminus U_{k_i}} | X_{U_{k_{i+1}}^\mathsf{c} \setminus U})\\
	\end{align*}
	where:
	\begin{inparaenum}[(a)]
		\item follows from the definition of a vertex;
		\item follows from the application of the inductive hypothesis;
		\item follows from the definition of conditional mutual information;
		\item \(U_{k'_1}^\mathsf{c} \setminus U' = U_{k_2}^\mathsf{c} \setminus U'\);
		\item \(U' \subseteq U_{k_1}^\mathsf{c}\) so partition \(U_{k_1}^\mathsf{c}\) into \(U_{k_1}^\mathsf{c} \setminus U'\) and \(U'\);
		\item follows from the chain rule for conditional entropy;
		\item follows from a change of variable for the sum index;
		\item follows from expressing the conditional mutual information in terms of the original set, and;
		\item follows from moving the first conditional mutual information into the sum.
	\end{inparaenum}
\end{IEEEproof}

\subsubsection{Full Proof of \propref{active-constraint}}
\label{sec:active-constraint-proof}
\begin{IEEEproof}
	\begin{equation}
		\begin{aligned}
			0 &\leq R(U) - H(X_U | X_{U^\mathsf{c}})\\
			  &= H(X_{U} | X_{U_{k_1}^\mathsf{c} \setminus U}) - H(X_U | X_{U^\mathsf{c}})\\
			  &+ \sum_{j = 2}^{m} I(X_{U \setminus U_{k_{j-1}}} ; X_{U_{k_j-1} \setminus U_{k_{j-1}}} | X_{U_{k_j}^\mathsf{c} \setminus U})\\
			  &= H(X_{U} | X_{U_{k_1}^\mathsf{c} \setminus U}) - H(X_U | X_{U_{k_1 - 1}}, X_{U_{k_1}^\mathsf{c} \setminus U})\\
			  &+ \sum_{j = 2}^{m} I(X_{U \setminus U_{k_{j-1}}} ; X_{U_{k_j-1} \setminus U_{k_{j-1}}} | X_{U_{k_j}^\mathsf{c} \setminus U})\\
			  &= I(X_{U}; X_{U_{k_1 - 1}} | X_{U_{k_1}^\mathsf{c} \setminus U})\\
			  &+ \sum_{j = 2}^{m} I(X_{U \setminus U_{k_{j-1}}} ; X_{U_{k_j-1} \setminus U_{k_{j-1}}} | X_{U_{k_j}^\mathsf{c} \setminus U})\\
			  &= \sum_{j = 1}^{m} I(X_{U \setminus U_{k_{j-1}}} ; X_{U_{k_j-1} \setminus U_{k_{j-1}}} | X_{U_{k_{j}}^\mathsf{c} \setminus U}).
		\end{aligned}
	\end{equation}
\end{IEEEproof}

\subsubsection{Proof of \propref{conditional-independence}}
\label{sec:conditional-independence-proof}
\begin{IEEEproof}
	Denote the elements of \(S\) ordered by \(\pi\) as \(t_1, \ldots, t_{|T|}, u_1, \ldots, u_{|U|}, v_1, \ldots, v_{|V|}, w_1, \ldots, w_{|W|}, y_1, \ldots, y_{|Y|}\).
	Then the elements of \(S\) ordered by \(\pi'\) is \(t_1, \ldots, t_{|T|}, v_1, \ldots, v_{|V|}, u_1, \ldots, u_{|U|}, w_1, \ldots, w_{|W|}, y_1, \ldots, y_{|Y|}\).
	We can show the following
	\begin{equation}
		\begin{aligned}
			R_\pi(t_i) &= H(X_{t_i} \mid X_{t_{i+1}}, \ldots, X_{t_{|T|}}, X_U, X_V, X_W, X_Y)\\
				   &= H(X_{t_i} \mid X_{t_{i+1}}, \ldots, X_{t_{|T|}}, X_V, X_U, X_W, X_Y)\\
				   &= R_{\pi'}(t_i)
		\end{aligned}
	\end{equation}
	\begin{equation}
		\begin{aligned}
			R_\pi(u_i) &= H(X_{t_i} \mid X_{u_{i+1}}, \ldots, X_{u_{|U|}}, X_V, X_W, X_Y)\\
				   &= H(X_{t_i} \mid X_{u_{i+1}}, \ldots, X_{u_{|U|}}, X_W, X_Y)\\
				   &= R_{\pi'}(u_i)
		\end{aligned}
	\end{equation}
	\begin{equation}
		\begin{aligned}
			R_\pi(v_i) &= H(X_{v_i} \mid X_{v_{i+1}}, \ldots, X_{v_{|V|}}, X_W, X_Y)\\
				   &= H(X_{v_i} \mid X_{v_{i+1}}, \ldots, X_{v_{|V|}}, X_U, X_W, X_Y)\\
				   &= R_{\pi'}(v_i)
		\end{aligned}
	\end{equation}
	\begin{equation}
		\begin{aligned}
			R_\pi(w_i) &= H(X_{w_i} \mid X_{w_{i+1}}, \ldots, X_{w_{|W|}}, X_Y)\\
				   &= R_{\pi'}(w_i)
		\end{aligned}
	\end{equation}
	\begin{equation}
		\begin{aligned}
			R_\pi(y_i) &= H(X_{y_i} \mid X_{y_{i+1}}, \ldots, X_{w_{|Y|}})\\
				   &= R_{\pi'}(Y_i)
		\end{aligned}
	\end{equation}
\end{IEEEproof}

\subsubsection{Proof of \thmref{alternative-conj}}
\label{sec:alternative-conj-proof}
The next lemma establishes that a convex combination of rates can be supported by a convex combination of supporting flows.
\begin{lem}
	\label{lem:convex-flow}
	Suppose \(R_i \in Q_{\sigma{SW}} \cap P_{\rho_c}\) and let \(f_i\) be a flow that supports \(R_i\).
	If \(R_\lambda = \sum_{i}\lambda_i R_i\) for \(\lambda_i \geq 0\) and \(\sum_{i}\lambda_i = 1\) then \(f_\lambda = \sum_{i}\lambda_i f_i\) is a flow that supports \(R_\lambda\).
\end{lem}
\begin{IEEEproof}
	Omitted for brevity.
\end{IEEEproof}

This is a restatement of and proof of \thmref{alternative-conj}.
\begin{thm}{Alternative to Conjecture}
	Let \(f^*_{R_i}\) be a min-cost flow that supports rate \(R_i\). Let \(R = \sum_i \lambda_i R_i\).
	The flow \(f = \sum_i \lambda_i f^*_{R_i}\) is a flow that supports \(R\) of minimum cost if there exists a vectors \((x^*(a) : a \in A)\) and \((z^*(v) : v \in V)\) such that for all \(i\)
	\begin{subequations}
		\begin{gather}
		x^*(a)(f^*_{R_i}(a) - c(a)) = 0\\
		(k(a) - x^*(a) - (z^*(\head(a)) - z^*(\tail(a))))f^*_{R_i}(a) = 0
		\end{gather}
	\end{subequations}
	for all \(a \in A\).
\end{thm}
\begin{IEEEproof}
	Having fixed a rate vector \(R_i\), we can solve for the min-cost flow for that rate with following LP
	\begin{equation}
	\label{eq:flow-primal}
	\begin{aligned}
		& \underset{f \geq 0}{\text{minimize}} & & \sum_{a \in A}k(a)f(a) & &\\
		& \text{subject to} & & f(a) \leq c(a) & & a \in A\\
		& & & f(\delta^{in}(v)) - f(\delta^{out}(v)) = 0 & &v \in N\\
		& & & f(\delta^{in}(v)) - f(\delta^{out}(s)) = -R_i(s) & & s \in S
	\end{aligned}
	\end{equation}
	and its corresponding dual
	\begin{equation}
		\label{eq:flow-dual}
		\begin{aligned}
			& \underset{x \leq 0, z}{\text{maximize}} & & \sum_{a \in A}c(a)x(a) - \sum_{s \in S}R_i(s)z(s) & &\\
			& \text{subject to} & & x(a) + z(\head(a)) - z(\tail(a)) \leq k(a) & & a \in A.
		\end{aligned}
	\end{equation}
	If \(R_i\) is a feasible rate vector, then there exists a min-cost flow \(f^*_{R_i}\) for this \(R_i\) and therefore optimal dual variables \((x^*_{R_i}, z^*_{R_i})\).
	Observe that the set of feasible dual variables does not depend on the rates \(R_i\), only on the edge costs \(k\).
	By assumption \(x^*_{R_i} = x^*\) and \(z^*_{R_i} = z^*\) for all \(i\) and therefore \((x^*, z^*)\) is dual feasible for \(R\).
	We have that by \lemref{convex-flow}, that \(f\) is primal feasible.
	Checking the complimentary slackness conditions for \(f\), \(x^*\), and \(z^*\), we have
	\begin{equation}
		\begin{aligned}
			x^*(a)(f(a) - c(a)) &= x^*(a)\left(\sum_i \lambda_i f^*_{R_i}(a) - c(a)\right)\\
					    &= \sum_i \lambda_i \left(x^*(a)(f^*_{R_i}(a) - c(a))\right)\\
					    &= 0
		\end{aligned}
	\end{equation}
	and similarly
	\begin{equation}
		(k(a) - x^*(a) - (z^*(\head(a)) - z^*(\tail(a))))f^*_{R_i}(a) = 0.
	\end{equation}
	We conclude that \(f\) is primal optimal and \(x^*, z^*\) are dual optimal solutions for a min-cost flow that supports \(R\).
\end{IEEEproof}

\subsubsection{Proof of \thmref{equal-path-costs}}
\label{sec:equal-path-costs-proof}
\begin{IEEEproof}
	Define \(\mu(v) \triangleq \min_{P} k(P)\) be the value of a min-cost \(v-t\) path in the network and let \(v \rightsquigarrow t\) indicate a \(v-t\) path.
	Let \(\hat{x}(a) = 0\) for all \(a \in A\) and \(\hat{z} = -\mu(v)\) for all \(v \in V \setminus \{t\}\).
	At \(\hat{x}, \hat{z}\), the constraints of \eq{dual} are equivalent to
	\begin{equation}
		\mu(\head(a)) + k(a) \geq \mu(\tail(a)).
	\end{equation}
	Observe that \(\tail(a) \rightarrow \head(a) \rightsquigarrow t\) is a directed \(\tail(a)-t\) path of cost \(\mu(\head(a)) + k(a)\) and therefore \(\hat{x}, \hat{z}\) is dual feasible.
	Furthermore, for every \(u \in V \setminus \{t\}\), there exists \(v \in V\) such that \((u, v) \in A\) and \(\mu(u) = \mu(v) + k((u,v))\).
	This means that there are at least \(|A| + |V| - 1\) \emph{active} constraints at \(\hat{x}, \hat{z}\) and it is a vertex.
	In fact, for a given \(v \in V\) all \(v-t\) paths have the same cost, \emph{all} constraints are active and \(\hat{x}, \hat{z}\) is the \emph{only} vertex of the dual feasible set.
	The dual feasible set is identical for all choices of source rates \(R\) and therefore the optimal solution is give by \(x^* = \hat{x}\) and \(z^* = \hat{z}\).
\end{IEEEproof}

\subsubsection{Proof of \thmref{optimal-check}}
\label{sec:optimal-check-proof}
\begin{IEEEproof}
	Ordering the elements of \(S\) according to the permutation \(\pi\) induces a nested family of subsets \(U_i \triangleq \left\{s_j : j \in [i]\right\}\).
	We construct a dual feasible \(y\) by setting \(y_U = 0\) for \(U\) not in the nested family and
	\begin{equation}
		\label{eq:subset-dual}
		y_{U_i} =
		\begin{dcases}
			\bar{h}(s_i) - \bar{h}(s_{i+1}) & i \in [n - 1]\\
			\bar{h}(s_i) & i = n.
		\end{dcases}
	\end{equation}
	We construct a dual feasible \(x\) from \eq{arc-dual}.
	Having primal feasible \((f^*_\pi, R_\pi)\) and dual feasible \((x, y, z)\), optimality follows from complimentary slackness.
\end{IEEEproof}

\subsubsection{Proof of \thmref{sigma-face-multi-sink}}
\label{sec:sigma-face-multi-sink-proof}
\begin{IEEEproof}
	The condition \eq{sigma-face-multi-sink} implies \(B_{\sigma_{SW}} \subseteq P_{\rho^{(t)}_c}\) for all \(t \in T\), which means that for any vertex of \(Q_{\sigma_{SW}}\) and any sink \(t \in T\) there exists a supporting virtual flow \(f^{(t)}\).
	If on the other hand \eq{sigma-face-multi-sink} does not hold, this implies there exists a sink \(t \in T\) and a vertex of \(Q_{\sigma_{SW}}\) for which no supporting flow \(f^{(t)}\) exists.
\end{IEEEproof}

\subsubsection{Proof of \thmref{multi-sink-bounds}}
\label{sec:multi-sink-bounds-proof}
\begin{IEEEproof}
	Denote an optimal solution to \eq{multi-sink-primal} as \((f^*, p^*, R^*)\).
	The optimal value of a min-cost flow for the single sink \(t\) is given by
	\begin{equation}
		\label{eq:lower-bound-sink}
		\sum_{a \in A}k(a)\hat{f}^{(t)}(a)
	\end{equation}
	and this must be a lower bound for the multi-sink problem with \(t \in T\).
	Suppose that it was not; then, 
	\begin{equation}
		\begin{aligned}
			\sum_{a \in A}k(a) \hat{f}^{(t)}(a) &\geq \sum_{a \in A}k(a) p(a)\\
							    &\geq \sum_{a \in A}k(a) f^{*(t)}(a)
		\end{aligned}
	\end{equation}
	which follows from the constraints of \eq{multi-sink-primal}.
	Note that the virtual flow \(f^{*(t)}(a)\) is a feasible solution to the single sink problem for sink \(t\), contradicting the assumption of \(\hat{f}^{(t)}\) as an optimal solution to the single sink problem for \(t\).
	Since \eq{lower-bound-sink} is a lower bound for every \(t \in T\), it must be true for
	\begin{equation}
		\label{eq:lower-bound}
		\max_{t \in T} \sum_{a \in A} k(a)\hat{f}^{(t)}(a).
	\end{equation}

	To produce an upper bound for the optimal value of \eq{multi-sink-primal}, we form a feasible solution from \(\left\{(\hat{f}^{(t)}, \hat{R}^{(t)}) : t \in T\right\}\) the set of optimal solution to single-sink problems:
	\begin{subequations}
		\begin{align}
			f^{(t)}(a) &= f^{*(t)}(a) \; \forall a \in A, t \in T\\
			R^{(t)}(s) &= R^{*(t)}(s) \; \forall s \in S, t \in T\\
			p(a) &= \max_{t} f^{(t)}(a).
		\end{align}
	\end{subequations}
	Finally, the cost of this feasible solution is
	\begin{equation}
		\sum_{a \in A}k(a) \max_{t \in T} \hat{f}^{(t)}(a).
	\end{equation}
\end{IEEEproof}
}{}
\end{document}